\documentclass[prx,twocolumn,amsfonts,showpacs]{revtex4-1}

\pdfoutput=1

\usepackage{epsfig}
\usepackage{psfrag}
\usepackage{amsmath}
\usepackage{amssymb}
\usepackage{color}
\usepackage{hyperref}
\usepackage{appendix}
\usepackage{booktabs}
\usepackage{url}

\begin{document}

\author{Wei-Can Yang$^{1,2}$}
\author{Chuan-Yin Xia$^{2}$}
\author{Hua-Bi Zeng$^{2}$}\email{ hbzeng@yzu.edu.cn}
\author{Makoto Tsubota$^{1,3}$}\email{tsubota@omu.ac.jp}
\author{Jan Zaanen$^{4}$}\email{jan@lorentz.leidenuniv.nl}

\affiliation{$^1$ Department of Physics, Osaka Metropolitan University, 3-3-138 Sugimoto, 558-8585 Osaka, Japan}
\affiliation{$^2$ Center for Gravitation and Cosmology, College of Physical Science and Technology, Yangzhou University, Yangzhou 225009, China}
\affiliation{$^3$ Nambu Yoichiro Institute of Theoretical and Experimental Physics (NITEP), Osaka Metropolitan University, 3-3-138 Sugimoto, Sumiyoshi-ku, Osaka 558-8585, Japan}
\affiliation{$^4$ Institute Lorentz for Theoretical Physics, Leiden University, Leiden, The Netherlands}

\title{Motion of a superfluid vortex according to holographic quantum dissipation.}

\begin{abstract}
Vortices are topological defects associated with superfluids and superconductors, which, when mobile, dissipate energy destroying the dissipation-less nature of the superfluid. 
The nature of this "quantum dissipation" is rooted in the quantum physical nature of the problem, which has been subject of an extensive literature. However, this has mostly be focused on the measures applicable in weakly interacting systems wherein they are tractable via conventional methods. Recently it became
possible to address such dynamical quantum thermalization problems in very strongly interacting
systems using the holographic duality discovered in string theory, mapping the quantum problem
on a gravitational problem in one higher dimension, having as benefit offering a more general view
on how dissipation emerges from such intricate quantum physical circumstances. We study here
the elementary problem of a single vortex in two space dimensions, set in motion by a sudden
quench in the background superflow formed in a finite density ”Reissner-Nordstrom” holographic
superconductor. This reveals a number of surprising outcomes addressing questions of principle.
By fitting the trajectories unambiguously to the Hall-Vinen-Iordanskii phenomenological equation
of motion we find that these are characterized by a large inertial mass at low temperature that
however diminishes upon raising temperature. For a weak drive the drag is found to increase when
temperature is lowered which reveals a simple shear drag associated with the viscous metallic vortex
cores, supplemented by a conventional normal fluid component at higher temperatures. For a strong
drive we discover a novel dynamical phenomenon: the core of the vortex deforms accompanied by a
large increase of the drag force.
\end{abstract}

\maketitle

\section{Introduction}

Macroscopic reality is characterized by dissipation, the fact that work is converted into heat under the governance of the second law of thermodynamics. In a recent
era the case has been fortified that this is a ramification
of many-body quantum physics. This is best expressed by the notion of eigenstate thermalization, wherein the stochastic nature of thermal physics is a corollary of the collapse of the wave function. Upon attaining the expectation values, the principles of irreversible thermodynamics emerge from the otherwise unitary quantum dynamics \cite{Alesio}. In fact, this quantum thermalization notion was already adapted earlier in the triumphant account of the finite temperature properties of the conventional quantum liquids in the 1950's: the superfluids/superconductors and Fermi-liquids \cite{AGD}. Considering that the nature of the (zero-temperature) quantum states differ from that encountered in normal ("molecular") fluids, the
thermal behaviour of the quantum fluids is accordingly
modified. In general, the computation of these properties is challenging owing to the hardships of non-equilibrium quantum many body physics. This classic development exploited the loop hole that physically relevant cases behave similar to a dilute gas that can be addressed using quantum kinetic Boltzmann theory; that is,  the Fermi-liquid and its BCS superfluid "derivative" as of relevance to $^3$He and conventional metals/superconductors, as well as the weakly interacting Bose gas  described by Bogoliubov theory and physically realized in the cold-atom laboratories.

A particular intricate affair in this context is the quantum thermalization behind the dissipative dynamics of {\em vortices}, which are the topological excitations of the superfluid. Their free movement causes the superfluid to transform into a normal dissipative fluid. This has been intensely studied in the context of the vortex fluids formed in external magnetic fields in condensed matter systems \cite{Blatter}. However, it was particularly scrutinized in the context of "quantum turbulence" by strongly "stirring" the superfluids formed from helium or cold atoms to turn them into a non-equilibrium tangle of vortex lines. It was found that their dynamical evolutions were reminiscent of hydrodynamical turbulence \cite{Vinen,Vinen2,Donnelly1,Finne}.

The derivation of dissipative dynamical properties of vortices from "first principle" quantum thermalization is a particularly challenging affair, which has been debated for decades \cite{Popov,Baym,Chandler,Duan,Kopnin,Thouless, Thompson, Simula, Thompsonthesis}. Inspired by the cold atoms, studies focused on the Bogoliubov Bose gas, wherein a certain degree of consensus was reached. However, the concept of Iordanskii force (the Magnus force exerted by the finite temperature final normal fluid component) remains unresolved \cite{Thompson,Thompsonthesis}.

Moreover, this main stream effort remains limited to "gaseous" circumstances, lacking methods to deal with strongly interacting quantum systems \cite{Penckwitt, Duine}. However, recent development indicated that holographic duality, or Anti-de Sitter/Conformal Field Theory (AdS/CFT) correspondence, as discovered in the context of quantum gravity in the string theory community \cite{Maldacena, Gubser1, Witten}, can be used as a highly flexible and accurate tool to compute quantum thermalization to handle strongly interacting circumstances that may be of relevance to condensed matter systems \cite{Zaanen1, Ammon, Lucas}.

This pertains to circumstances that are in a manner opposite to the conventional gas limit. It is by now well understood that the quantum physics described by holography is associated with extremely strongly coupled and densely entangled forms of matter \cite{Zaanen2,Zaanen3} that are characterized for instance by extremely rapid thermalization physics \cite{OTOC}. Such a quantum system is, through holographic duality, mapped on a classical gravitational physics in one higher dimension that may be enumerated by solving the Einstein equations revolving around black holes in the gravitational bulk encoding for the finite temperature physics \cite{Witten,Danielsson}. A remarkable development is the "fluid-gravity duality" \cite{Bhattacharyya,Hubeny, Rangamani,Balm}, where holography has been used to reconstruct the hydrodynamical theory describing the finite temperature macroscopic physics of such strongly coupled quantum fluids. This addressed intricate issues such as the structure of higher gradient hydrodynamics and the radius of convergence of the hydrodynamic expansion.

Superfluidity is part of this agenda in the form of the holographic superfluids \cite{Gubser2,Hartnoll,Herzog}, as based on this, the two-fluid phenomenology has been reconstructed in great detail \cite{Herzog2,Yarom}. However, under certain technical restrictions, it is comparatively easier to study the dynamical time evolution of moving vortices, wherein their energy is transferred to heat. This was already explored in pioneering works addressing superfluid turbulence in 2D \cite{Chesler,yutian,Du,Wittmer,Chuanyin,Yang,Yang2,Chesler2,Lan, Ewerz} and studies of the reconnection dynamics of vortex lines in 3D using holography \cite{Wittmer2}. Here we will focus  on the most realistic holographic superconductor that can be handled presently in this context: that is, the finite density Reissner-Nordstrom (RN)  holographic superconductor.

This propagates the study of holographic vortices a step further compared to this state of the art that revolves around the "minimal" holographic superconductors \cite{Chesler,yutian,Du,Wittmer,Chuanyin,Yang,Yang2,Chesler2,Lan, Ewerz,Wittmer2} being only a good proxy of physical systems at higher temperatures close to $T_c$. The RN superconductor is characterized by its own pathology -- the underlying metallic state with its zero temperature entropy, e.g. \cite{Zaanen3}. In addition, considering the current computational capabilities, it is still impossible to consider the dynamical modification of the bulk geometry in the presence of the moving vortex, which is the gravitational backreaction \cite{Chesler3, Hong}. However, this only becomes important at very low temperatures and the RN superconductors offer a window to study vortex dissipation of a sufficiently rich physical toy model over a large range of temperatures, which may not be necessarily associated with anything existing in the laboratory.

We focused on the most elementary dynamical issues. Consider a departure from a single vortex "pancake" at rest in the 2 dimensional finite temperature superfluid, to a sudden switch to a background superflow, and then to following the trajectory of the vortex in space and time. These trajectories contain all the information required to reconstruct the effective Hall-Vinen-Iordanskii (HVI) phenomenological equations of motion (EOM) \cite{Vinen,Iordanskii}. We determine its dissipative- and reactive coefficients and compare these with the standard Gross-Pitaevskii (GP) based analysis obtained using the kinetic theory \cite{Gross, Pitaevskii, Tsubota, Kobayashi3}. The earlier holographic studies of vortex dynamics \cite{Chesler,yutian,Du} were focused on the much more complicated physics associated with the manner in which the vortex tangles, that were not forced externally, evolved in time. This study can be considered as the ground work, considering the most elementary dynamical evolution, and thus shedding light on the ingredients involved in the many vortex context.

We find the outcomes of our study of this holographic
toy model revealing in the regard that it offers an alter-
native, broader view on the factors governing vortex dissipation as rooted in the microscopic physics. Within its limitations, it is a highly disciplined affair, strictly obeying underlying physical principles that are still computable given the power of holographic duality. This perspective, in certain crucial aspects, differs from and is generalized on the established wisdoms of the usual kinetic gas circumstances. In summary, these are:

\begin{itemize}
\item[a.] Is a vortex characterized by a finite inertial mass? This question has caused considerable confusion and debate in the past; however, the answer is exquisitely dependent on the underlying microscopy. The RN holographic vortex offers in this regard a vivid illustration, capturing the two extremes simultaneously. For well understood reasons, in a weakly coupled fermionic (BCS) superconductor, this mass becomes infinitesimal; whereas, a long time ago, it was already understood that in a dense, strongly interacting boson superfluid such as $^4$He, it is of order of the atom mass. Our holographic superconductor evolved smoothly from the former- to the latter extremes through the simple lowering of temperature from $T_c$ (Section \ref{massdet}), with the caveat that this mass was set by the chemical potential dealing with the ultrarelativistic finite density matter of holography.
\item[b.] According to the HVI EOM, the moving vortex dissipates its energy via a drag coefficient. According to the Bogoliubov theory this drag is nearly entirely caused by its motion relative to the normal fluid component referring to the two fluid phenomenology. In contrast to the "empty" cores of the bosonic vortices, the RN vortex cores are "filled" with a "strange metal" fluid characterized by the famous minimal viscosity. This is not different from the situation encountered in metallic BCS superconductors. We find that its damping was now dominated by a simple viscous shear drag associated with the motion of this "droplet" of normal fluid in case of a weak drive (Section \ref{Dshear}). However, employing a highly
sensitive holographic method revealing the origin
of the total energy dissipation we also identify a
component that is like in the boson case associated
with the normal fluid (Section \ref{totaldis}).
\item[c.] The vortex was set in motion by the Magnus force exerted by the superfluid; however, there remains room for such a force owing to the normal fluid component; that is, the "Iordanskii" force. The status of this force has been subject of considerable confusion even in case of the weakly interacting bosons. It involves an influential claim that it should disappear altogether. The Iordanskii force can be determined with a high precision in our set up, and it was found to be finite and characterized by a surprising temperature independence in the small drive regime (Section \ref{Iordanskiiforce})
 \item[d.] Further, the flexibility of holography facilitated a closer inspection into the strong non-equilibrium regime wherein the strength of the drive was of order of the intrinsic scales of the system. However, the question is what happens with the vortex motion when the background superfluid starts to approach the (Landau) critical velocity, when considering the present set up. Here, an entertaining new physics was obtained. In the strongly coupled case,  the internal structure of the vortex was {\em deforming} in nature. It behaved similar to a speeding boat, thus developing a "bow wave" at its front and a "wake" at its back (Section \ref{vortexwake}). Consequently, a strong effect was exerted on the drag, which increased by several orders of magnitude in case of low temperatures (Section \ref{deformeddiss} ).

\end{itemize}

The remainder of this paper is organized as follows. To render this presentation as self-contained as possible both for the holographic- and vortex community, the various ingredients used were reviewed in the next Section (\ref{methods}). In addition, a description of our holographic set up including the HVI equation and the Gross-Pitaevskii "industry standard" of the cold atom community used as comparison have been presented. The bulk of the paper as described in the above will then
unfold, and we finish with a concluding section putting
our findings in a broader perspective

\section{The methodology: holography, Gross-Pitaevskii and the Iordanskii-Vinen-Hall equation of motion.}
\label{methods}

This section presents the various computational- and fitting methods used in this study. This revolves in the first place around the holographic
description of vortex motion.  The setup used -- minimal holographic superfluidity in the Reissner-Nordstrom background -- is a standard one, as is the manner of numerically solving the dynamical bulk EOM's employed in this study. Thereafter, the Gross-Pitaevskii theory that was used as a template familiar to the study of particularly cold atoms as a comparison is summarized. This was used to compare the behaviour of the holographic vortices. Finally, this section elucidates the effective equation of motion governing the dynamics of vortices: the HVI equation. Experts in quantum turbulence will not find anything
new in this part of this Section: this serves in first
instance to render this exposition to be self contained
also for the non-experts in vortex dynamics.

\subsection{Reissner-Nordstrom holographic superfluidity.}
\label{modeldetails}

The discovery of holographic superconductivity in 2008 \cite{Gubser2, Hartnoll,Herzog} has crucial in triggering the exploration of holography in condensed matter physics. This began with the discovery that the addition of a scalar to a simple Schwarzschild Anti-de-Sitter bulk geometry encoding for a zero density finite temperature boundary CFT violates the Breitenlohner-Friedman (BF) stability bound upon lowering the temperature. Consequently, the black hole acquires a scalar "hair," which is allowed by AdS asymptotes. A response occurs in the boundary in the absence of a source, signaling the spontaneous breaking of the $U(1)$ symmetry in the boundary.

This triggered an extensive follow up literature, exploring and systematically improving on this "minimal" construction. The first step was to depart from a finite density normal metallic state in its simplest incarnation, in the form of the dual to the simple charged black hole with Einstein-Maxwell gravity in the bulk. This was the "Reissner-Nordstrom" (RN) strange metal. Its space-time metric in Eddington-Finkelstein coordinates was $ds^2=\frac{L}{z^2}[-f(z)dt^2-2dtdz+dx_i^2]$, where $f(z)=1+Q^2(z/z_h)^4-(1+Q^2)(z/z_h)^3$ and $z$ is the (extra) radial dimension of the bulk. Further, $Q=\frac{\mu z_h}{2}$ is the characteristic parameter associated with the chemical potential $\mu$ and the Hawking temperature is expressed as $T=(3-Q^2)/4\pi$, thus setting the temperature in the boundary \cite{Romans,Chamblin,Liu}. The extremal case with $Q^2=3$ was the dual to the zero temperature boundary, that has been intensely studied in the early era of the condensed matter applications \cite{Zaanen1, Ammon, Lucas}.

An attractive feature is that the boundary is characterized by a diverging dynamical exponent $z$, indicative of the (quasi) local quantum criticality that appears to be observed in the strange metals realized in cuprate high Tc superconductors \cite{Zaanen2,Zaanen3}. However, it is also pathologically characterized by a finite horizon area at am extremum, implying a temperature independent entropy for $T/\mu << 1$ that is finite at zero temperature. Similar to all holographic fluids, its macroscopic behaviour at finite temperature is governed by hydrodynamics characterized by the "minimal viscosity" $\eta = 1 / (4\pi) (\hbar/k_B) s$ where $s$ is the entropy density. It becoming temperature independent as well at low temperature. A crucial aspect is that such a fluid resides {\em inside} the vortex cores formed in this finite density holographic superfluid.

To realize such a boundary condensate, a complex scalar field $\Psi$ is added to the bulk Einstein-Maxwell action, which is expressed as,

\begin{align}
S=\frac{1}{16\pi G_N} \int d^4x \sqrt{-g}\Big[\mathcal{L}_{EH}-\mathcal{L}_{matter}
\Big]
\\
\mathcal{L}_{EH}=R+6/L^2
\\
\mathcal{L}_{matter}=\frac{1}{4}F_{\mu\nu}F^{\mu\nu}+|\partial_\mu \Psi-iqA_\mu \Psi|^2+V(|\Psi|^2)
\end{align}

where $\mathcal{L}_{EH}=R+6/L^2$ is the Einstein-Hilbert action with cosmological constant $6/L^2$ wiring in the asymptotic AdS geometry. Further, $F_{\mu\nu}$ is the Maxwell field strength with vector potential $A_{\mu}$, that is coupled minimally to the scalar involving the charge $q$. Pending the form of the potential $V$, when the temperature is lowered, the BF bound of the scalar may be violated such that the hair emerges; according to the dictionary, the complex scalar field $\Psi({\bf r},z)$ is dual to a complex scalar operator $\psi({\bf r})$ representing the superfluid order parameter on the boundary.

The bulk field forms asymptotes near the boundary as $\Psi = \phi z + \psi z^2 + \mathcal{O}(z^3)$ whereas, the gauge field $A_M({\bf r},z)$ is dual to a conserved $U(1)$ current, forms asymptotes such as $A_\mu =a_\mu +b_\mu z + \mathcal{O}(z^2)$. In the presence of the hair, in case of $\psi({\bf r})$, a response in the absence of a source $\phi({\bf r})$ was observed. This signals the condensation of the complex scalar. In the superfluid regime $a_x=a_y=0$ were set as the Dirichlet boundary conditions for $A_x$ and $A_y$ at $z=0$ \cite{Domenech}. It is well established
that this reconstructs impeccably the phenomenology of
the superfluid state, including the Landau order param-
eter theory but also the two-fluid phenomenology on the
linear response level.

Close to the transition temperature the geometry con-
tinues to be the RN-AdS one in the presence of the
hair, but upon lowering temperature it starts to back
react on this geometry. At zero temperature the horizon disappears completely, and depending upon the details of the scalar potential and the charge $q$, various other geometries may be realized \cite{Gouteraux}.
 However, handling of the non-stationary and inhomogeneous vortices remains technically impossible owing to difficulties in keeping track of this gravitational back reaction and we are forced to ignore it.

Technically, this can be rigorously imposed by considering the large $q$ limit that suppresses the backreaction. However, these backreaction effects become important for large $q$ only at the lowest temperatures that we will avoid. Another aspect is that generically holographic superfluid reconstructs the Landau-Tisza two fluid hydrodynamical phenomenology, characterized by a normal fluid component  in addition to the superfluid with a density $\rho_n (T)$ that decreases with reduction in temperature. When ignoring the backreaction, the normal fluid behaves as though it is pinned (motionless), acting as a perfect heat bath.

\begin{figure}[h]
\includegraphics[trim=4.5cm 10.5cm 0cm 9cm, clip=true, scale=0.75, angle=0]{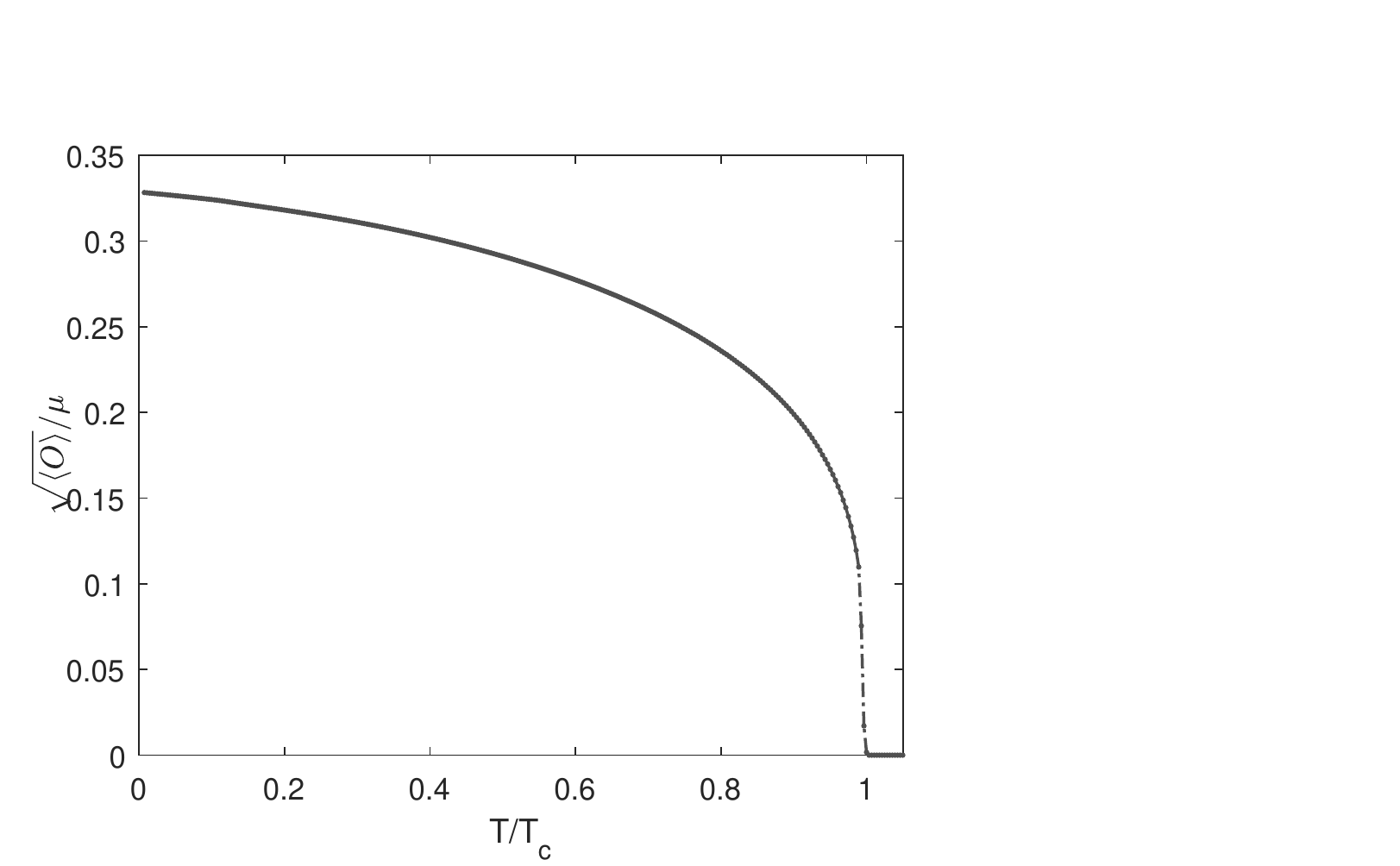}
\caption{Order parameter evolution as a function of temperature, characterizing the "minimal" Reissner-Nordstrom holographic superconductor while ignoring the gravitational backreaction. Close to the critical temperature, it shows the characteristic $\sim (1-T/T_c) ^{\beta}$ with critical exponent $\beta=1/2$, the behaviour associated with the Landau mean-field.}\label{orderparameter}
\end{figure}

This study expressed the potential $V$ considering the "minimal" form $V(|\Psi|^2)=m^2|\Psi|^2$ with $m^2=-2$. The outcome for the temperature evolution of the boundary order for this potential was in the form shown in Fig. (\ref{orderparameter}). As we already stressed it encodes for the two fluid phenomenolog, while the normal fluid density $\rho_n \sim T^{1.26}$ \cite{Arean} is well below $T_c$.

 \begin{figure}[h]
\includegraphics[trim=4.5cm 9cm 0cm 9cm, clip=true, scale=0.7, angle=0]{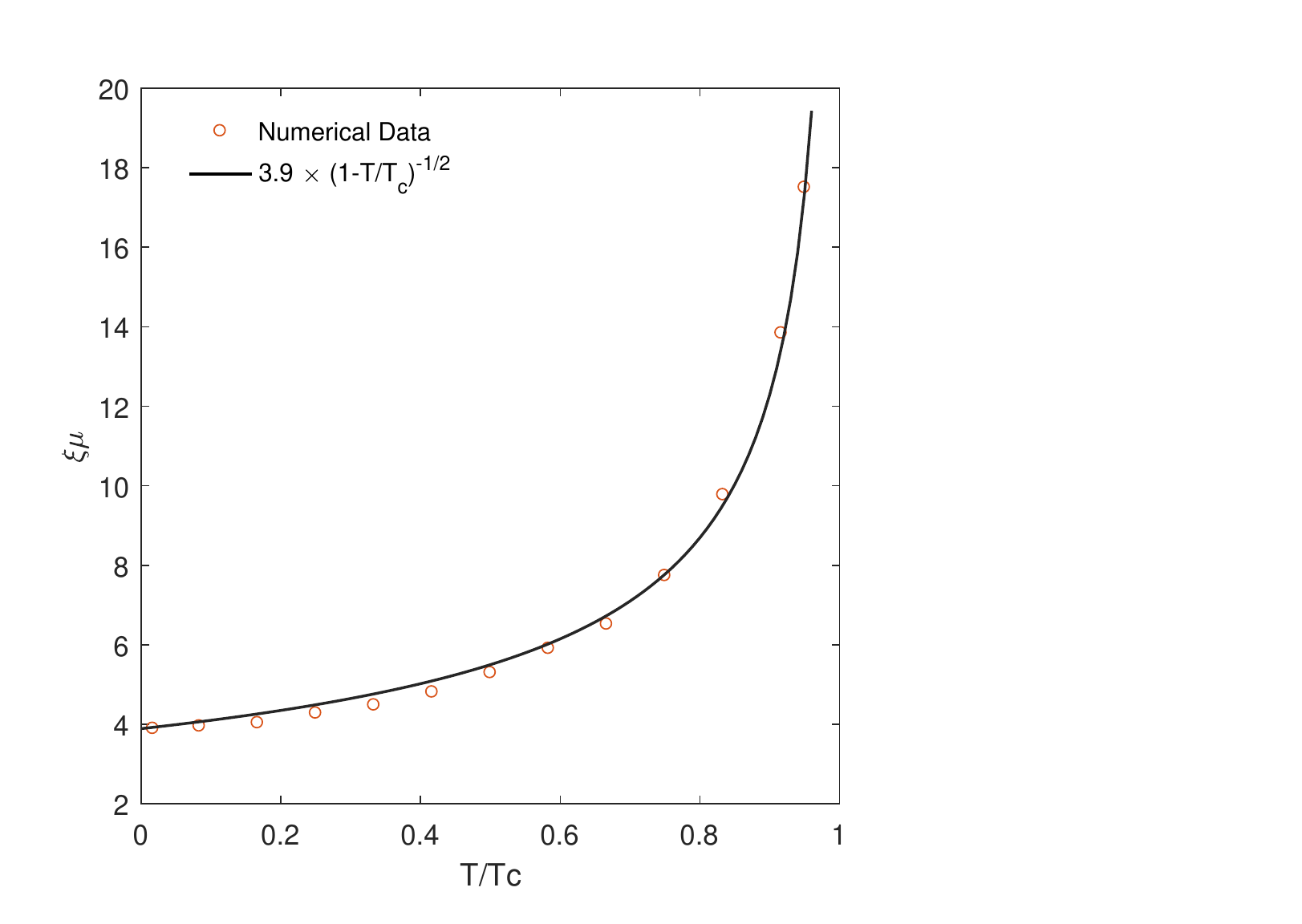}
\caption{Radius of the core of the static vortex corresponding with the coherence length in units of $1/\mu$ as a function of temperature for the minimal RN holographic superconductor. The red points are the numerical results and the blue lines is the fitted curve that is precisely obtained according to the Landau mean-field expectation. Notice that at low temperature this indicates a strongly coupled "local pair" similar to a superconductor as $\xi \rightarrow 4/\mu$.}\label{size}
\end{figure}

Vortices are the ubiquitous topological excitations of a complex scalar order parameter and these were thoroughly explored in holographic superconductivity and superfluidity \cite{Montull, Dias, Tallarita, Maeda, Albash, Keranen, Domenech, Bao, Bao2}. A crucial property is that their cores "punch a hole" in the scalar hair, which exposes the black hole horizon; that is. holographic vortices are characterized by a "strange" metallic core that is dissipative. This core dissipation was identified as a crucial factor by Chesler et al. \cite{Chesler}. It was claimed that it was the reversal of the 'inverted cascade" of 2D hydrodynamical turbulence when entering the superfluid state. These "metallic cores" will be focused upon in the remainder of this study.

Holographic superconductivity is associated with pairing as a BCS superconductor. A property of this particular set up is that it converts at a {\em low} temperature in a "strongly coupled" superconductor similar to a BCS-BEC cross over; that is, the coherence length ("pair size") shrinks to microscopic dimensions and at low temperature it can be considered as a dense, strongly interacting "local pair" bosonic system (like $^4$He) \cite{Zaanen1,Zaanen3, QCBCS, SchmalianholoSF}. This can be illustrated by the temperature evolution of static vortices. Their core size is set by the coherence length and Fig. (\ref{size}) shows the computed outcome.

It conformed to the expectations of Landau mean field implying that $\xi \sim ( 1 - T/T_c)^{-1/2}$. However, the microscopic cut-off length was set by $\xi = 4/\mu$ and it was observed that at low temperatures it saturated at this value. This is consistent with the gap $\Delta$ to $T_c$ ratio being $2 \Delta/T_c \sim 8.4$ and was much larger than the weak coupling BCS value $\sim 3.4$, while $T_c \simeq \mu/4$. The chemical potential exhibited a similar role here as the Fermi energy \cite{Zaanen3} indicating that the pairing scale was of the order of the Femi-degeneracy scale. However, upon raising temperature, $\xi$ increases, and close to $T_c$ it becomes like a weakly coupled BCS superconductor. We will find that this reflects in a very interesting manner in the temperature dependence of the vortex inertial mass (Section \ref{massdet}).

Upon ignoring the backreaction, full non-equilibrium, dynamical evolution of the vortex system in the boundary can be easily tracked by numerically solving the remaining equations of motion in the bulk system. The bulk equilibrium geometry the EOM of the bulk gauge- and scalar fields can be written as

\begin{eqnarray}
d_\nu F^{\mu \nu}=J^\mu=iq(\Psi^\star D^\mu \Psi - \Psi D^\mu \Psi^\star)
\\
D_\mu D^\mu \Psi -m^2 \Psi=0
\end{eqnarray}

Upon imposing the holographic boundary conditions, these can be solved numerically. Mainly, high order Runga-Kutta methods following Ref. \cite{Chesler} were employed. This has proven to be considerably powerful, even capable of handling significantly complex situations such as tracking the quantum turbulence evolution of an initially dense vortex tangle that is annealing away as function of time. Here, a single two (space) dimensional "vortex pancake" was focused upon. We prepared the initial static vortex structure using the superposition phase method, wherein a phase factor $\mathrm{exp}(i\phi)=\mathrm{exp}(i \mathrm{arctan}[(y-y_i)/(x-x_i)])$ was multiplied with the global bulk scalar field $\Psi(z)=\psi(z) e^{i\phi}$ to evolve it over time, and finally annealing into a stationary vortex \cite{Han}. Subsequently, an otherwise homogeneous superflow was imposed on the boundary by assigning a finite value to the bulk $a_y$ field that was dual to the supercurrent, to track the time evolution of the system.

\subsection{Reference: the Gross-Pitaevskii equation of the cold atoms.}
\label{GPreview}

In particular, when dealing with the weakly interacting Bose-Einstein gas realized in cold atom systems, there appears to be a community consensus
that the non-equilibrium time evolutions, such as our vortex-acceleration set up, are governed by the effective Gross-Pitaevskii (GP) equation of motion \cite{Reeves, Kobayashi, Tsubota, Kobayashi2,Kobayashi3,Penckwitt, Duine}. Thus, this study attempted to reproduce the expectations of the motion of such vortex according to this consensus in the particular dynamical setting considered. This was implemented for comparison purposes.

The GP equation describes damped motion associated with a relaxing Landau order parameter. Departing from the order parameter ($\psi$) free energy of form $- \mu
|\psi|^2 + g | \psi |^4$ where $\mu \sim (1 - T/T_c)^{1/2}$, the EOM is postulated to be,

\begin{equation}
 (i-\gamma)\hbar \partial_t \psi({\bf r},t) = [-\frac{\hbar^2}{2m}\nabla^2 +g|\psi({\bf r},t)|^2-\mu - {\bf u \cdot p}]\psi({\bf r},t)
 \label{GPE}
\end{equation}

The crucial assertion that is not self-evident is that the dissipation can be captured by a simple dissipative parameter $\gamma$ damping the order parameter. A substantial literature has evolved, aimed at determining this parameter based on microscopics, while considering the quantum kinetic gas theory. This is then associated with the particle exchange with the normal fluid that acts a heat bath and it was found that $\gamma = \frac{6 a^2}{\pi q^2} k_B T$ \cite{Penckwitt, Duine,Kobayashi2,Kobayashi3}, where $a$ is the s-wave scattering length of the atoms and $q$ is a time-independent variational parameter.

In addition, the temperature evolution of the superfluid density is required as an input in our set up. However, this is easy considering the Landau mean-field theory underneath, such that the superfluid density $\rho_s = |\psi |^2 = \mu / (2 g)$. The highest dissipation parameter that can be achieved was selected as the critical temperature. Through comparisons of different temperature, this study obtained the parameters of the HVI equation through the vortex motion.
The outcomes for the dissipative parameters governing the vortex motion (HVI equation) are shown in Fig. \ref{smallvsD} and these are usually interpreted as reflecting the temperature dependence of the normal fluid density that is responsible for the damping of the vortex motion in the weakly interacting boson gas.

The GP dynamical system was studied following a similar methodology as in case of the holographic system. The superposition phase method was used to generate a single vortex pancake to then quench an external background superflow ${u \cdot \bf p}={\bf u}\cdot (-i\hbar \nabla)$, where ${\bf u}$ is the velocity. Subsequently, the system was integrated as a function of time.

\subsection{Forcing a vortex: the parameters of the Hall-Vinen-Iordanskii (HVI) equation.}
\label{HVIequation}

A vortex in isolation is characterized by a conserved topological charge and it behaves similar to a particle in two dimensions, or a string in three dimensions. The question of the manner in which such an object reacts to external- and internal forces is therefore well defined. Based on general considerations Hall, Vinen, and Iordanskii \cite{Vinen,Iordanskii} reported a generic equation of motion.

In three dimensions, vortices form lines, and for every line element a normal velocity component is defined. For simplicity, this study specialized to two dimensions, where the vortex was simply a "particle" moving with a velocity ${\bf v}_L$. The physics being considered employed of the same principle in case of 2D and 3D.

This "HVI" equation is highly phenomenological although quite general. In fact, it appears that the only crucial assumption for its validity is that the moving vortex behaves as a {\em rigid} object lacking internal degrees of freedom. As a novelty revealed by the present study, Section (\ref{deformation}) shows that according to holography, there exists a transient regime that opens up at strong drives even when this was not the case and the vortex was actually {\em deforming}. However, it was also found that at the short times where the vortex began to accelerate (Section \ref{massdet}) and the longer time drag dominated steady flow regime (Sections \ref{dissipation}, \ref{deformeddiss}) this HVI equation could accurately fit the computed trajectories even under these strong non-equilibrium circumstances.

The point of departure of the HVI equation is the two-fluid phenomenology: the vortex coexists and interacts both with the superfluid- and the normal fluid components characterized by their velocities ${\bf v}_s$ and ${\bf v}_n$ and temperature dependent densities $\rho_s$ and $\rho_n$, respectively. In principle, the dissipationless superflow exerts a purely reactive {\em Magnus} force ${\bf F}^{(S)}$ on the vortex. This is proportional to the difference ${\bf v}_L - { \bf v}_s$, with a magnitude set by the quantum of circulation ${\bf k}=\hat{\bf z} h / m$ (in non-relativistic units) and the superfluid density $\rho_s$. However, the normal component may also exert such a Magnus force with a strength set by the unknown coefficient $D'$ (the "Iordanskii force"). In addition, there should be a dissipative drag force that is only proportional to the velocity difference ${\bf v}_n- {\bf v}_L$ and a strength set by the coefficient $D$. Finally, the vortex may have an inertial mass $M_v$ and combining these aspects results in the HVI equation of motion for the vortex particle,

\begin{eqnarray}
M_v\frac{d{\bf v}_L}{dt} & = & {\bf F}^{(S)}+{\bf F}^{(N)} \nonumber \\
{\bf F}^{(S)} & = & \rho_s({\bf v}_s- {\bf v}_L)\times {\bf k} \nonumber \\
{\bf F}^{(N)} & = & - D' ({\bf v}_n- {\bf v}_L)\times {\bf k} + D( {\bf v}_n- {\bf v}_L)
\label{IordanskiiEq}
\end{eqnarray}

This equation has been subjected to intense experimental and theoretical study as it was reported a long time ago; for example, Ref.'s \cite{Thompson, Donnelly2, Kopnin2, Kobayashi, Fedichev, Bevan, Simula,Thouless, Wei, Fedichev}.

As a simplifying circumstance, both in GP and the holographic case the normal fluid is by construction always at a standstill. This implies that ${\bf v}_n = 0$. Consequently, the EOM is further simplified into a form that was used to fit the vortex trajectories:

\begin{eqnarray}
M_v\frac{d{\bf v}_L}{dt} & = & {\bf F}^{(S)}+{\bf F}^{(N)} \nonumber \\
{\bf F}^{(S)} & = & \rho_s({\bf v}_s- {\bf v}_L)\times {\bf k} \nonumber \\
{\bf F}^{(N)} & = & D' {\bf v}_L \times {\bf k} - D {\bf v}_L
\label{IordanskiiEqvn0}
\end{eqnarray}

Our study is unique with regard to employing a {\em minimal} dynamical protocol. We depart from a single vortex pancake annealed to equilibrium. At time $t=0$, the background superfluid was made to flow with a velocity $v_s$ in the $y$ direction. According to the $F^{(S)}$ term, this superflow exerted a "side wise" Magnus force on the vortex; and assuming its inertial mass was finite it began to accelerate in the $x$ direction. This motion was damped by the drag $\sim D$ and at long times the vortex entered a stationary flow regime. Consequently, the HVI equations were converted to those for EOM, for the $x$ and $y$ components of the vortex velocity,

\begin{eqnarray}
   M_v \frac{dv_{Lx}}{dt} &  = & \rho_skv_{s}-(\rho_s-D')kv_{Ly}-Dv_{Lx} \nonumber \\
    M_v \frac{dv_{Ly}}{dt} & =  & (\rho_s-D')kv_{Lx}-Dv_{Ly}
\label{HVIquench}
\end{eqnarray}

This particular "current quench" set up employed in this study was convenient for the extraction of the parameters from the actual trajectories were numerically computed. The next Section focuses on the determination of the inertial mass in the short time regime. Departing from $v_L =0$, the initial trajectory will merely reflect the accelerating vortex in the $x$-direction where the Magnus force is exerted according to

\begin{equation}
    M_v \frac{dv_{Lx}}{dt}=ma_x=\rho_skv_{s} \label{masseq}
\end{equation}

However, in the drag dominated stationary regime at long times, $D$ and $D'$ can be directly deduced from the "terminal" $x$ and $y$ velocity components of the vortex according to Eqs (\ref{HVIsteady},\ref{HVIsteadyvel}).

\begin{figure}[h]
\includegraphics[trim=1cm 8cm 0cm 7.5cm, clip=true, scale=0.5, angle=0]{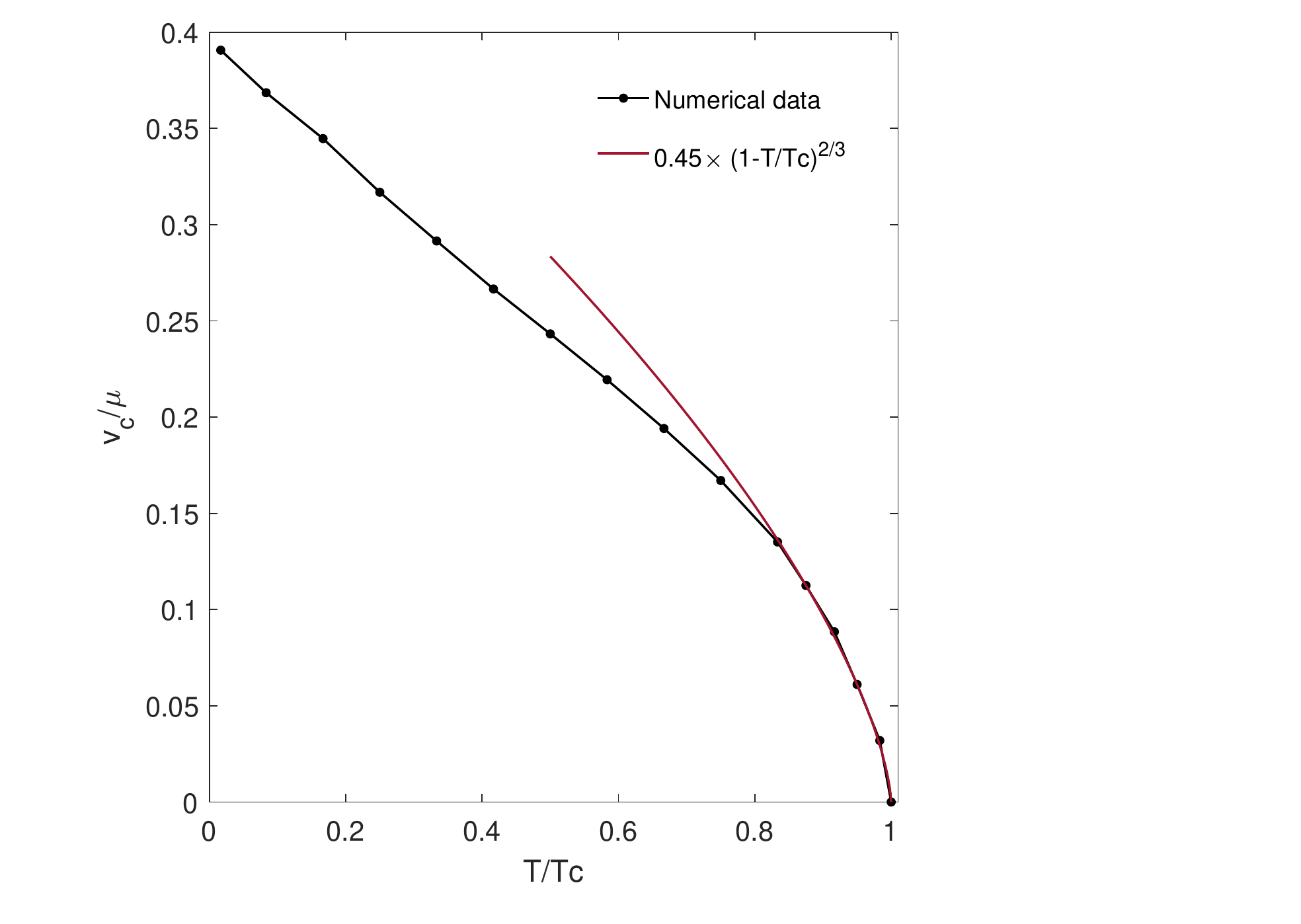}
\caption{Plot of the critical velocity $v_c$ of the holographic superfluid versus temperature. Near the critical temperature, it satisfies the mean-field relation $v_c \sim (1-T/T_c)^{2/3}$. Away from the critical
Regime near $T_c$, it crosses over to $\sim (1-T/T_c)$ upon lowering temperature, to saturate at $\sim 40$ \% of the natural velocity scale set by $\mu$.}\label{vc}
\end{figure}

\section{The short time regime and the vortex mass.}
\label{massdet}

Let us now turn to the results of our holographic sim-
ulations, contrasting them with the community-standard
GP results. Once again, at $t=0$ a superflow was switched on in the $y$ direction characterized by a velocity $v_s$. As inferred previously, this exerted a Magnus force on the vortex and the non-equilibrium realms were entered into. However, this implies yet another control parameter: the degree of non-equilibrium, which is dependent on the magnitude of $v_s$. However, the question is the manner in which this influence can be normalized.

The relevant quantity is the Landau-Tisza critical velocity. For an increasing superflow the superfluid order parameter decreases to diminish completely at the critical velocity $v_c$, where the system turns normal. It is well known how to determine this for holographic
superconductor \cite{Herzog}: the critical velocity of the homogeneous case is determined by the point where the order parameter disappears, and the outcome is shown in Fig. (\ref{vc}). This shows a reasonable behaviour that is similar as
to what is found in conventional superfluids: the critical
velocity tracks the order parameter to a degree, vanishing at $T_c$ to increase in a mean-field fashion as in experiment according to $v_c \sim (1-T/T_c)^{2/3}$ \cite{James}. Upon further lowering of temperature it crossed over to a linear rise to reach a maximum at $T=0$ \cite{Varoquaux}

The magnitude of the drive with regard to where the real non-equilibrium phenomena is expected to appear is therefore associated with the ratio $v_s/v_c$. When this ratio is very small the "near equilibrium" regime will be explored, which is the subject of this section. However, when the ratio becomes "of order unity," strong non-equilibrium effects that appear as highlighted in Section(\ref{Sectdeformation}), should be considered. Moreover, these realms can always be entered into when close to $T_c$. This is because $v_c \rightarrow 0$. However, as a helpful circumstance, it was found that according to our holographic set up, such genuine dynamical phenomena requires low temperature (explained later).

\begin{figure}[h]
\includegraphics[trim=2.9cm 10cm 2cm 8.5cm, clip=true, scale=0.6, angle=0]{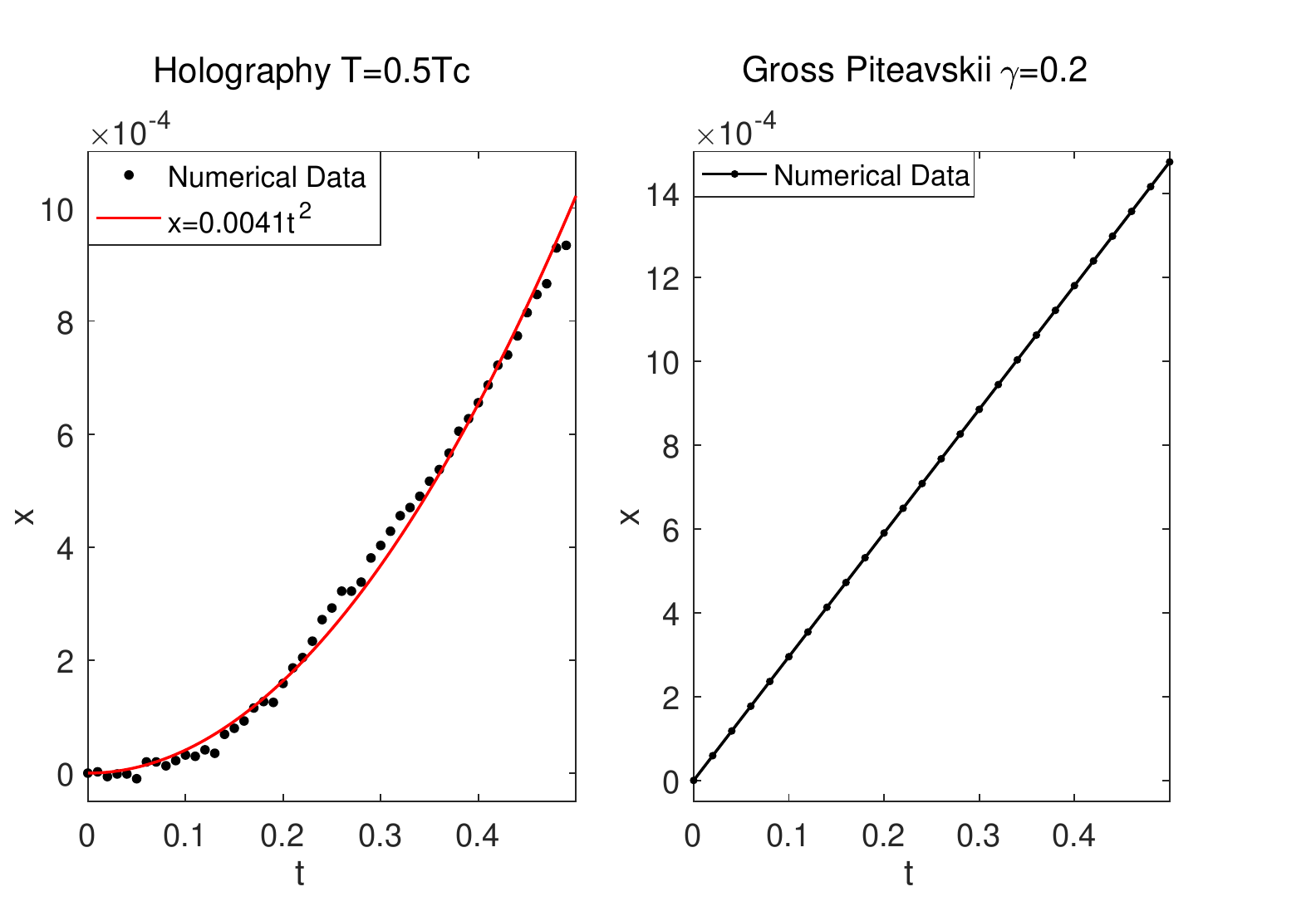}
\caption{The trajectory of vortex motion in x-direction perpendicular to the external superflow in the y-direction with a velocity $v_s / \mu = 0.0025$. In the left panel the holographic result is shown for $T = 0.5 T_c$ (blue dots) as well as the fit to the HVI equation of motion (red line); the curvature signals the finiteness of the inertial mass of the vortex. The right figure shows the result for the Gross-Piteavskii model with dissipation rate $\gamma=0.2$ showing the instantaneous acceleration associated with the vanishing mass hard wired in this construction. }\label{RNGP}
\end{figure}

\begin{figure}[h]
\includegraphics[trim=2.4cm 10cm 0cm 10cm, clip=true, scale=0.55, angle=0]{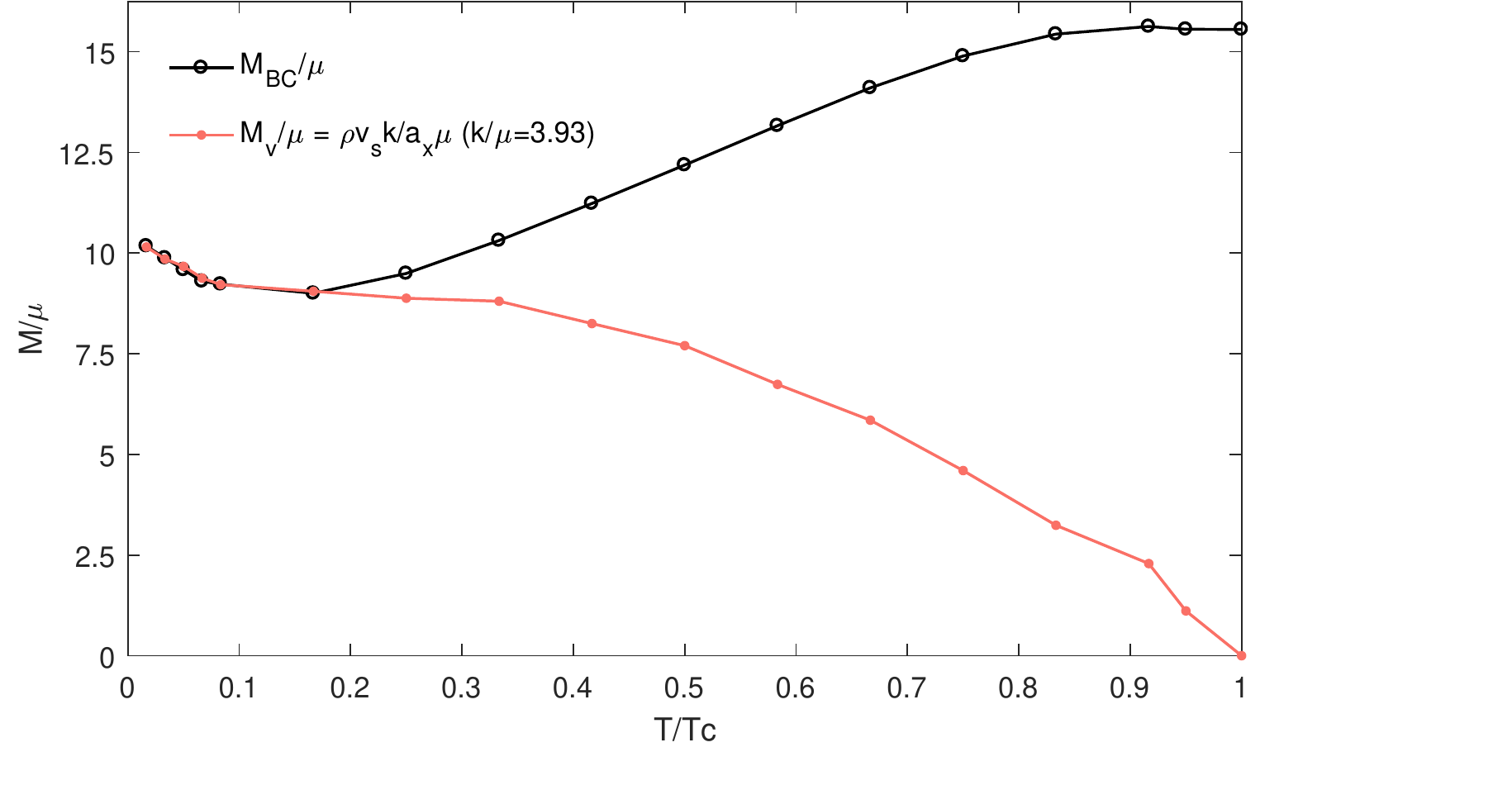}
\caption{Red line: the inertial mass of the holographic vortex in units of the chemical potential $\mu$ (see the main text) as function of temperature, determined by fitting its trajectories to the HVI equation as in Fig.(\ref{RNGP}). The comparison with the Baym-Chandler mass computed using Eq. (\ref{BCmass}) (black line) demonstrates that at low temperature the holographic superconductor is "local pair" like, crossing over to the vanishing mass BCS-like behaviour upon approaching $T_c$.
}\label{mass}
\end{figure}

As a reminder, according to the EOM Eq. (\ref{IordanskiiEqvn0}) a typical trajectory appears as follows. The Magnus force encoded by ${\bf F}^{(S)}$ will yield a force in the orthogonal $x$-direction of the form ${\bf F}^{M}=\rho_s{\bf v}_{sy}\times {\bf k}$ that will accelerate the vortex according to $M_v\frac{d{\bf v}_{Lx}}{dt}= {\bf F}^{M}$. At later times, the forces encoded by ${\bf F}^{(N)}$ will take over, both modifying the overall Magnus force and compensating for the drag. The vortex "pancake" is similar to a particle subjected to an external force and when its inertial mass is finite, it can be deduced from the short time trajectory according to Eq. (\ref{masseq} ).

A typical outcome is shown for the trajectories in Fig. (\ref{RNGP}). In the GP case (right panel), it is a straight line; however, this is of course expected. It is assumed at the onset that the vortex is devoid of inertial mass and the acceleration to the terminal velocity is instantaneous. This may appear to be unphysical; however, this is the actual mechanism in the "gaseous" weakly coupled (BCS, Bogoliubov) systems where vortices are extremely light. In contrast, an excellent fit of the trajectory was obtained using the HVI equation for the holographic case. It can be discerned from the figure that a curvature of the trajectory at short times $x \sim t^2$ signaled a finite inertial mass that also affected the trajectory at longer times in a manner consistent with a finite mass Iordanskii equation.

To determine this mass in absolute units, the manner in which the various quantities are defined when dealing with the (ultra) relativistic must be inspected, with zero rest mass matter determining the UV of the holographic system. According to Eq. (\ref{masseq}), its mass follows from the initial time evolution as $M_v=k\frac{\rho_s v_s}{a_x}$, where $k$ is the quantized circulation of the vortex superflow. In a system characterized by a mass $m$, the quantum of circulation is universally set by ${\bf k}={\bf \hat{z}}h/m$; however, the remains with regard to what can be considered for ${\bf k}$, when the rest mass $m$ is zero. In this finite density holographic system, the only scale is the chemical potential.

The quantum of circulation is of course also a universal topological number in this relativistic fluid, and it can be deduced as follows. In the non-relativistic setting, $k=h/m$ is rooted in the dimensional factor transforming the gradient of the phase into velocity $v_{\phi}=\frac{\hbar}{m}\nabla \phi$,

\begin{equation}
    \oint l dv_{\phi}=\frac{\hbar}{m}\oint dl \nabla \phi = \frac{2\pi\hbar}{m}n=\frac{h}{m}n
\end{equation}

The phase velocity of our holographic superconductor has been determined by Herzog and Yarom \cite{Herzog2, Yarom} in their computation of “fourth sound”,

\begin{equation}
    v^2_{\phi}=\frac{\rho_s}{\mu \frac{\partial \rho_s}{\partial \mu}}c^2
\end{equation}

At low temperature $\rho_s \simeq \rho$, whereas for a conformal system $\rho \sim \mu^{d-q}$ ($d$ is the boundary spacetime dimension). This implies in explicit units that

\begin{equation}
    v_{\phi}=\frac{1}{\sqrt{d-1}}\frac{\hbar c^2}{\mu}\nabla \phi
\end{equation}

Here, $\#_{\phi}=2\pi/(\mu\sqrt{d-1})$ is defined and natural units are used to yield

\begin{equation}
    \oint dlv_{\phi}=\#_{\phi}\oint dl\nabla\phi=\frac{2\pi}{\mu\sqrt{d-1}}n
\end{equation}

It can be concluded that the {\em universal} value of the circulation quantum for our massless $2+1D$ holographic superconductor was set by the chemical potential as $k/\mu=\sqrt{2}\pi=4.44$. Coupled with the availability of $\rho_s$ in numerical form, the mass can now be determined in units of $\mu$. This absolute mass as obtained by fitting the trajectories to the HVI equation is plotted as function of temperature in Fig. (\ref{mass}).

This revealed a surprise: the inertial mass of the holographic vortex was strongly temperature dependent, being large in an absolute sense at low temperatures, to decreased with increase in temperature and vanished upon approaching $T_c$, where it behaved like a GP vortex in this regard!

In fact, this issue of the mass of the vortex has been subjected extensive debates. Contradicting claims have appeared, with one extreme case reported by Baym and Chandler \cite{Baym} for this mass in $^4$He. A dense bosonic superfluid, set by $M_{BC}=\frac{1}{2}\rho\pi\xi^2$ was considered, where $\rho$ is the mass density and $\xi$ the coherence length setting the size of the vortex core. This relied on the correct assertion that in such a superfluid, the $^4$He atoms were expelled from the vortex core implying $M_{BC}$ to be of order of the He atom mass, that is, being "large".

The other extreme is found in the weakly coupled BCS theory for fermionic superconductivity and the Bogoliubov bosons of the GP equation (e.g., \cite{Kopnin}) where the mass is predicted to be very small. The resolution is actually well known. This mass is a UV sensitive affair that is well understood in case of conventional BCS type superconductors. This involves the ratio of the BCS gap $\Delta$ and the Fermi energy $E_F$. In the weak coupling regime, only a fraction $\simeq \Delta/E_F$ of the fermions forming the Fermi sea formed Cooper pairs. Consequently, the coherence length (pair size) was very large. This involves only states near $E_F$, characterized by an energy independent density of states signaling an emergent {\em charge conjugation symmetry}. This implies that the electronic density stays {\em homogeneous} in the presence of a vortex. Matter is not accumulated in the vortex core and therefore its mass will vanish.

However, upon increasing the BCS coupling the gap will eventually become of order of the Fermi energy and the coherence length becomes of order of the interparticle distance, which are the "local" or "preformed" pair regime. It is widely believed that the cuprate high Tc superconductors may approach this regime. The physics in this regime is reminiscent of a dense bosonic superfluid such as $^4$He, with the "local pairs" functioning as the bosons. in the most important in this context is that for $\Delta$ of order of $E_F$, the {\em charge conjugation asymmetry} is manifested. Consequently, the vortex core is derived from matter as in $^4$He and its inertial mass becomes "large." This motive was crucial in the context of the rather strongly coupled cuprate high $T_c$ superconductors in terms of model considerations involving the simple attractive U fermion Hubbard model \cite{Feiner}. Moreover, it has been claimed that particular magnetotransport anomalies in the type II Abrikosov lattice phase of cuprates are owing to such a net electrical charge associated with the vortex cores\cite{Khomskii}.

 As we already emphasized in Section (\ref{modeldetails}), our holographic superconductor generally behaves as a fermionic, BCS like superconductor with the chemical potential $\mu$ taking the role of the Fermi-energy. Simultaneously, in the large $N$ limit, the thermal order parameter fluctuations are suppressed and the superconducting order parameter that is tied to a gap scale, as in BCS, behaves in a Landau mean field manner as a function of temperature (Section (\ref{modeldetails})). Perhaps best illustrated by the coherence length (Fig. \ref{size}), the holographic $\Delta / \mu << 1$ near $T_c$ and the ensuing effective charge conjugation causes the vortex mass to vanish. However, upon lowering the temperature, this ratio is rapidly increased to reach a "local pair" like limit at low temperature as we already highlighted in Section (\ref{modeldetails}). Thus, we conclude that in this holographic superconductor, the whole vortex mass repertoire is realized, starting from a vanishing to a large mass, trough simple temperature variation.

The degree of closeness of the vortex approach to the local pair limit can be examined by estimating the appropriate Baym-Chandler mass. In this relativistic setting, this mass is associated with the internal field energy \cite{Popov,Duan} $E_{field} = M_{BC} c^2$ where $E_{field} = \mu \Delta \rho$ and $\Delta \rho$ is according to Baym and Chandler associated with the density deficit at the vortex core associated with the expulsion of matter from the core,

\begin{equation}
  \Delta \rho  =  \int (\rho_s-\rho({\bf r}))d{\bf r},
\label{BCmass}
\end{equation}

where $\rho({\bf r})$ is the radial distribution of the density in the presence of the vortex as function of the radial coordinate $r$ and $\rho_s$ is the uniform density. Owing to complete access to the profile $\rho({\bf r})$, this can be precisely determined, and the outcome is shown In Fig. (\ref{mass}) in the form of the black line. As evident, at temperatures below $\simeq 0.2 T_c$, this estimate coincided precisely with the holographic computation.

In fact, a slight upturn was observed at the lowest temperatures in both cases. This was associated with the fact that the decrease in the area $\sim \xi^2$ slightly overcompensated the increase in the density contrast. This is a testimony of the precision of the Baym-Chandler estimate at low temperature.

The Baym-Chandler vortex mass estimate can also be determined at higher temperatures, as characterized by the growth of the coherence length/vortex size. The result, as shown in Fig. (\ref{mass}), indicates that it is fairly temperature independent; it increases slightly, maintain a large value up to $T_c$. This is in strong contrast with the case of the holographic mass, wherein the mass decreased smoothly and vanished entirely upon approaching $T_c$. However, as emphasized, this is according to the expectation for the "BCS-like" fermionic pairing underlying the holographic superconductivity.

\section{Dissipating the vortex motion: holography versus Gross-Piteavskii for a small drive.}
\label{dissipation}

Next, consider the dissipative aspects associated with the vortex motion. As emphasized in the introduction, this involves quantum thermalization, which is a tall order in general for conventional approaches. Much of the established tradition has focused on the weakly interacting Bose gas as captured by Bogoliubov theory, being tractable through quantum kinetic theory. This is also directly applicable to the cold atom condensates, and was used as a comparison to highlight the contrast between this community consensus and the holographic outcomes. This study focused on near-equilibrium, considering the vortex motion for small $v_s$. As shown later, in this small drive regime, the HVI equation works very well capturing the vortex motions. Subsequently, the dissipation is, in the first instance, captured by the vortex "drag" quantity $D$. In addition, at a finite temperature the $D'$ Iordanskii force parameter quantifying the Magnus force associated with the normal fluid must be considered. Although a reactive quantity, it requires finite temperature and is therefore also part of the thermalization agenda.

\begin{figure}[h]
\includegraphics[trim=2cm 7cm 0cm 6.8cm, clip=true, scale=0.53, angle=0]{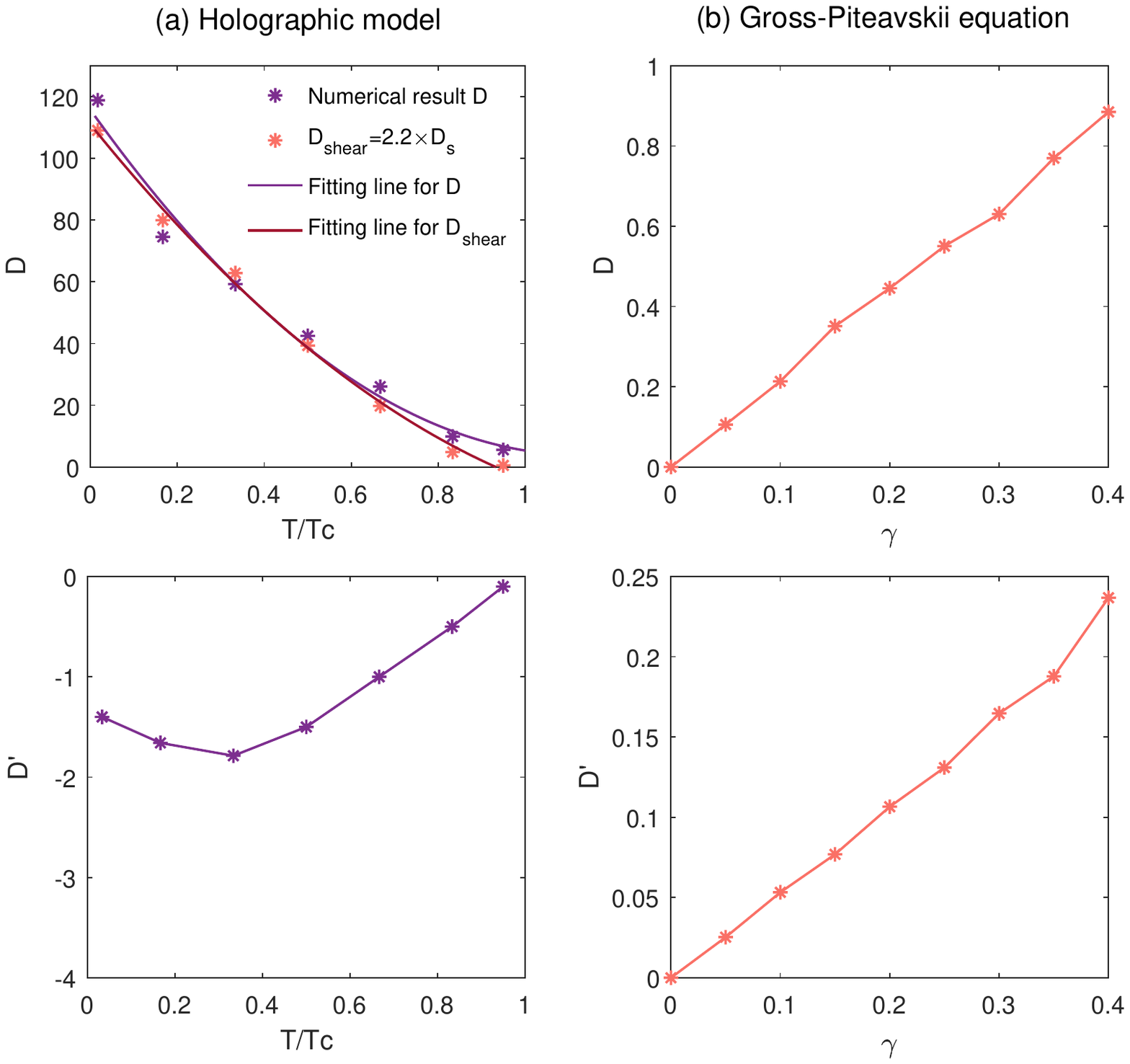}
\caption{Coefficient $D$ (upper panel) and $D'$ (bottom panel) of the HVI equation of motion for a vortex driven in a small external flow $v_s/\mu=0.0025$. The outcome for the holographic vortex is shown on the left, and for comparison the typical expectation following from the Gross-Pitaevskii modelling of the Bose gas is shown on the right. $D$ parametrizes the drag and in case of the holographic vortex, it {\em increases} upon lowering of temperature, which is in contrast with GP. As the comparison with the dimensional analysis estimate $D_{shear}$ shows, this is nearly entirely owing to a simple shear drag of the metallic core, while the small deviation at higher temperatures reveals the contribution of the normal fluid component (see main text).
}\label{smallvsD}
\end{figure}

In our set up, in the drag dominated stationary regime the vortex moved with a constant velocity in both the $x-$ and $y$ directions. As $d v_L/dt =0$, the HVI equation simplifies to

\begin{eqnarray}
\rho_s v_s k & =& (\rho_s - D') v_{L_y} k + D v_{L_x} \nonumber \\
0 & =&(\rho_s - D') v_{L_x} k + D v_{L_y}
\label{HVIsteady}
\end{eqnarray}

The first line expresses that the motion in the x direction is sourced by the superfluid moving in the y-direction. However, as it moved through the super- and normal fluid at rest, the combined Magnus force ($\rho_s -D'$) induced a velocity component $v_{Ly}$ in the supercurrent direction. This further added to the $v_x$ component (second line). Subsequently, both velocity components were subjected to a simple drag force $\sim D$. Solving Eq. (\ref{HVIsteady}) indicates that from the measured $v_x$ and $v_y$ velocities the unknown quantities $D,D'$ are expressed as

\begin{eqnarray}
D & =& \frac{v_s v_{L_x}}{v^2_L} \rho_s k \nonumber \\
D' & =& \left( 1-\frac{v_s v_{L_y}}{v^2_L} \right) \rho_s
 \label{HVIsteadyvel}
\end{eqnarray}

where $v^2_L = v^2_{Lx} + v^2_{Ly}$. As $v_s$, $k$ and the temperature dependent $\rho_s$, are known, the coefficients $D$ and $D'$ can therefore be determined in absolute magnitude through simple measurement of $\vec{v}_L$ at late times (Fig. \ref{smallvsD}).

\subsection{Drag parameter $D$ and the viscosity of the vortex core.}
\label{Dshear}

First, consider the GP comparison template. Using the parametrization introduced in Section (\ref{GPreview}), the temperature dependence of $D$ and $D'$ for the GP case as determined by fitting the vortex velocity is shown in the right panels of Fig. (\ref{smallvsD}). This indicates that $D \simeq T$ \cite{Fedichev}, and this temperature dependence is interpreted in the two fluid phenomenology in terms of the vortex experiencing a drag proportional to the normal fluid fraction $\rho_n$ \cite{Kobayashi,Kobayashi2,Kobayashi3,Iordanskii}. According to this parametrization the Iordanskii force parameter $D'$ also increases $\sim T$. This reflects Iordanskii's original proposal, indicating that this should be also set by $\rho_n$. However, this appears to be controversial to the present day, with various claims, including a prediction based on general principles, indicating that it should vanish \cite{Thompsonthesis,Thouless}.

Next, consider the holographic vortex. Figure (\ref{smallvsD}) shows that the temperature dependence of the drag parameter $D$ is entirely different from the GP case. Instead of being proportional to the normal fluid density $\rho_n$ that disappears at zero temperature (holographic system), $D$ actually {\em increased} upon lowering of the temperature. It reached a maximum value in case of $T=0$.

This simply reveals that the vortex drag was nearly completely dominated by the dissipation associated with the normal "strange metal" core. Moreover, this can be quantified in terms of an elementary dimensional analysis consideration. This reveals that to function as a conventional hydrodynamical shear-drag associated with the fact that the normal core behaves like a viscous fluid, while the spatially inhomogeneous nature of the vortex breaks the translational symmetry. It is similar to the simple Poiseuille flow problem wherein it was explained based on the resistance of a low Reynolds number fluid flowing through a pipe while the coherence length was the pipe diameter.

Consider the shear viscosity $\eta$ of the fluid. Using relativistic quantities, this can be converted into the transversal momentum diffusivity or kinematic viscosity by $D_{\pi} = \eta /(\epsilon +P)$, where $\epsilon$ and $P$ are the energy density and pressure ($\sim n m c^2$ in non-relativistic fluids), respectively. The dimension was $\left[ D_{\pi} \right] = m^2 /s$. In the presence of a characteristic length associated with the translational symmetry breaking, associated with the vortex size/coherence length $\xi$, a characteristic current relaxation rate $\Gamma = D_{\pi}/ \xi^2$ is obtained as $\left[ \Gamma \right] = 1 /s$

 Next, consider the HVI equation. Here, $\left[ D \right] = kg/s$ while the characteristic mass scale was associated with $M_v$. Hence, we conclude that by dimensional analysis,

\begin{equation}
D_{\mathrm{shear}} = A_{D} M_v \frac{\eta} {\epsilon + P} \frac{1}{\xi^2} =A_D D_s
\label{dragestimate}
\end{equation}

where $A_D$ is a dimensionless parameter of order unity.

In holographic systems including the RN metal, the viscosity is governed by the famous "minimal viscosity" formula,
$\eta = \frac{\hbar}{4 \pi} s$ where $s$ is the entropy density (we are using natural units, $\hbar =1$). Henceforth, $D_{shear}$ was entirely determined by thermodynamic factors ($s,\varepsilon, P$) which are available in closed form for the RN metal, and in addition the known quantities $\xi$ and $M_v$.

It was found that this $D_{\mathrm{shear}}$ accurately captured the temperature dependence of $D$. In Fig. (\ref{smallvsD}), the green line indicates the $D_{\mathrm{shear}}$, which owing to fitting yielded $A_{D} \simeq 2.2$. Future studies can attempt to investigate why this dimensionless parameter acquires this value. Qualitatively, the RN metal is famously pathological considering that in the low temperature regime, $s$ is approximately temperature independent and set by the zero temperature entropy. Similarly, $\epsilon + P \simeq \mu n + s T $, was also temperature independent \cite{Almeida,Iqbal,Sachdev}. The increase in $D$ with decrease in the temperature is therefore, in first instance, caused by the increase in the mass upon lowering $T$ (Fig \ref{mass}) combined with the decrease in the vortex size (Fig. \ref{size}), while also $\epsilon + P$ contributes.

\subsection{The core and the drag of the normal fluid: the total energy dissipation.}
\label{totaldis}

From the above the origin of the qualitative behaviour
of the drag at low temperature is clear.  However, the fit is not perfect and there is room for more. In fact, as for
the GP case the effective phenomenology described by
holography is coincident with the two fluid hydrodynamics. When temperature rises the normal fluid raises its
head and should contribute to the vortex drag as well.

\begin{figure}[h]
\includegraphics[trim=0.5cm 9cm 0cm 8cm, clip=true, scale=0.45, angle=0]{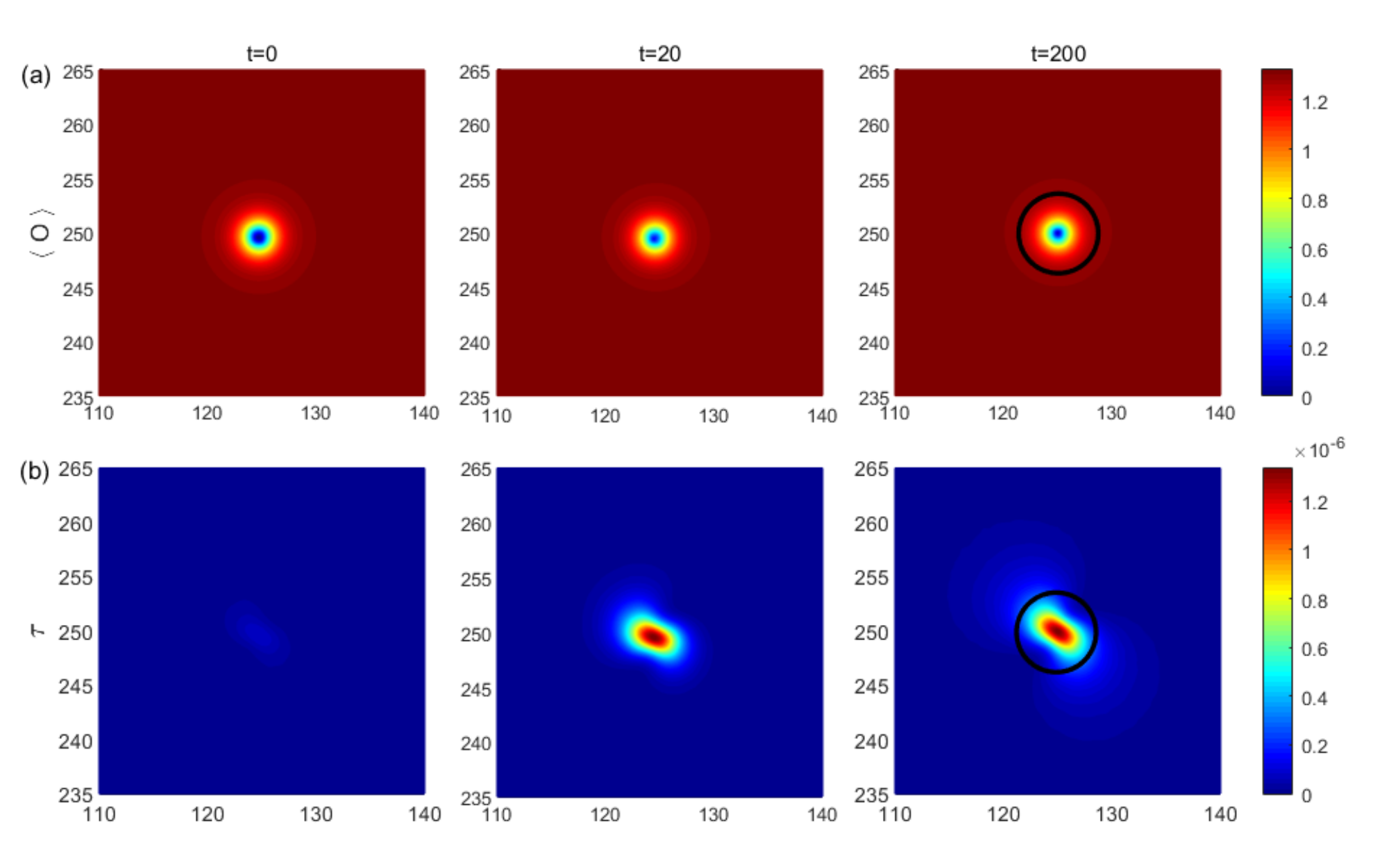}
\caption{Real space images of the time evolution of temperature $T=0.5T_c$ holographic vortex associated with a small velocity $v_s / \mu = 0.0025$ current quench. We depart from the initial state ($t = 0$, left), where in the middle panel the vortex is still accelerating ($t = 20$) and the right panel represents the long time stationary flow ($t=200$). The upper panels show the scalar field amplitude revealing the vortex core, with a radius indicated by the black circle in the right panel (e.g., Fig. \ref{size}). The lower panels show the total dissipative energy flux as determined in the bulk by the infall through the horizon. As in the lower right panel highlights, at this rather low temperature, this is dominated by the shear drag of the metallic core; however, there is a discernible contribution outside the core area associated with the drag originating from the normal fluid component.
 }\label{move}
\end{figure}

\begin{figure}[h]
\includegraphics[trim=3.2cm 7.5cm 2cm 6cm, clip=true, scale=0.65, angle=0]{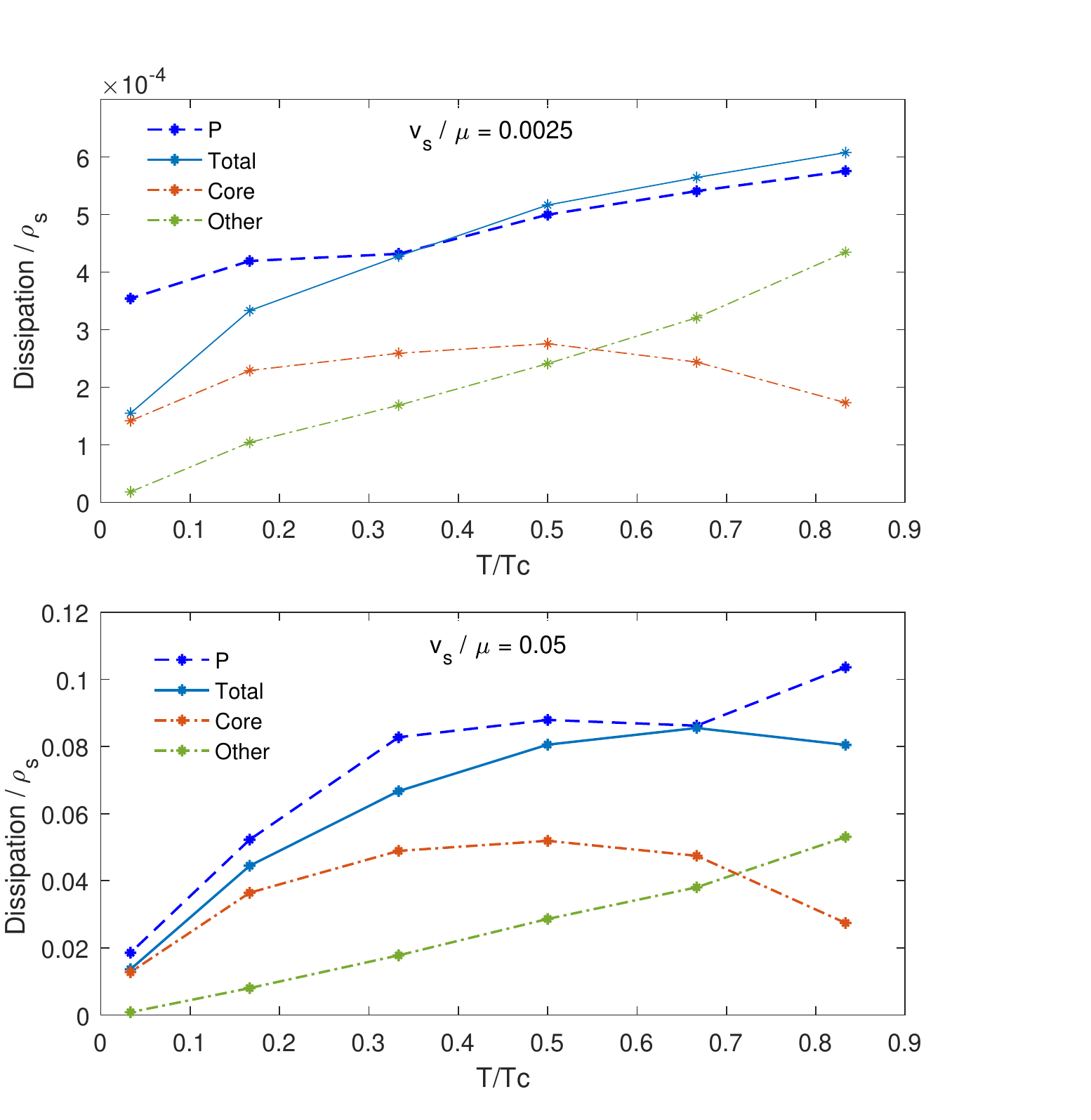}
\caption{ The energy flux through the bulk black hole horizon measures the dissipated power of the holographic vortex. The integrated "total" flux (full blue line) as a function of temperature for a small- ($v_s/\mu=0.0025$, upper panel) and large ($v_s/\mu=0.05$, lower panel) velocity is compared with the power dissipation implied by the HVI equation
(dashed blue line). Using this horizon measure, the dissipated power can be separated into core- (red dashed-dotted) and other (green dashed-dotted) contributions. This shows that the core contribution dominated at low temperature, whereas the non-local contribution associated with the normal fluid fraction was dominant at higher temperatures.
 }\label{dissipation}
\end{figure}

We can now exploit a particular flexibility of holog-
raphy to compute a property that is otherwise hard to
measure.  This facilitates the monitoring of the core- and normal fluid contributions to the vortex drag $D$ as a function of temperature.

It is a famous dictionary entry that the {\em total} energy dissipation in the boundary is governed by the falling of metric perturbations (gravitons) through the horizon in the bulk. The dissipative energy flux $\tau$ can be computed by considering the covariant conservation of the probe stress tensor $\mathcal{T}$ \cite{Chesler} with bulk EOM,

\begin{equation}
   \tau = \mathcal{T}^z_t|_{u=1}=\frac{1}{2}(F_{0i}F^{zi}+D_0\Psi^{\ast}D^{z}\Psi+D^z\Psi^{\ast}D_0\Psi)|_{u=1}
    \label{Energyflux}
\end{equation}

The flux $\mathcal{T}^z_t|_{u=1}$ through the bulk black hole horizon at $u=1$ integrated over the spatial manifold represents the total energy dissipation. Figure (\ref{move}) shows a typical result of the appearance of this energy dissipation for the moving vortex. Even at the relative high temperature of $T =0.5 T_c$, much of the energy dissipation is associated with the core. It "gravitons falling through the hole in the hair" can be observed directly. However, beyond the core area (black circle), there exists certain energy flux associated with the “GP style” normal fluid dissipation. Before considering this aspect, first, this sum-rule like quantity was used determine whether it matched the dissipation implied by the HVI equation. This represents a high precision test, wherein it checked if the HVI captures the motion of the holographic vortex.

For this purpose we focus on the long time stationary flow regime where the external Magnus forces exerted by the background superflow and the drag force were balanced. The dissipation rate is expressed as the work done by the external force times the velocity in the direction where this force is exerted. According to Eq. (\ref{IordanskiiEqvn0}),

\begin{eqnarray}
P={\bf F}^{(S)} \times {\bf v}_L & = & \rho_s({\bf v}_s- {\bf v}_L)\times {\bf k} \times {\bf v}_L \nonumber \\
&=& \rho_s {\bf v}_s \times {\bf k} \times {\bf v}_L \\
&=& F_{ext} v_{Lx}
 \label{Power1}
\end{eqnarray}

 Consequently, the strength of the external Magnus force can be defined according to $ F_{ext} = \rho_s k v_s$. Now, the steady state velocity must be determined in the x-direction, which can be obtained from Eq. (\ref{HVIsteady}). Through certain simple algebra,

\begin{equation}
 P = v_{Lx} F_{ext} = F_{ext}^2 \times \frac{D}{D^2+D'^2k^2-2D'k^2\rho_s+k^2\rho_s^2}
\label{Power2}
\end{equation}

Using the values for the HVI quantities $\rho_s, D, D'$as we
determined in the above from the trajectories (and ther-
modynamics) we can compute $P$ and compare it with the total horizon flux computed holographically using Eq.(\ref{Energyflux}). The outcome is shown in Fig. (\ref{dissipation}) both for a very small ($v_s = 2.5 \; 10^{-3} \mu$) and relatively large ($v_s = 5 \; 10^{-2} \mu$) drives. As evident, the total horizon flux and the HVI power dissipation $P$ (Eq. \ref{Power2}, dashed lines) were quite consistent; for the small drive the numbers became small at low temperature such that resolution was lost.

How does this power dissipation actually work? We observe that it {\em decreased} with the reduction in temperature; the HVI expression Eq. (\ref{Power2}) can be considered for an explanation. In a first step, $D'$ can be ignore because it is numerically very small, see underneath. Putting $D' = 0$,

\begin{equation}
P = \frac{F_{ext}^2}{D ( 1 +\frac{\rho_s^2k^2}{D^2})}
\label{Power3}
\end{equation}

This may look unfamiliar. Let is compare it with the
familiar Ohmic dissipation where the role of the external
force is taken by the bias $F_{ext} \rightarrow V$, the velocity was similar to the current $v_{Lx} \rightarrow I$, and the role of the drag coefficient fulfilled by the resistance $D \rightarrow R$. The Ohmic power dissipation $P = V \times I = V^2/ R \rightarrow F_{ext}^2/ D$. Compared with Eq. (\ref{Power3}), it can be inferred that this simple Ohmic type dissipation is recovered when $ \rho_s k << D$. The regime where $ rho_s k >> D$ exhibited the "anomalous" scaling $P \simeq D v^2_s$ was used instead. In reality, this is just a ramification of the fact that the Magnus force is set by the difference between the vortex- and "driving" velocity. For large $D$ this does not play a role and the dissipation is of the usual Ohmic type; whereas, for small $D$ it implies the seemingly anomalous "$P \sim R$" behaviour.

According to Fig. (\ref{smallvsD}), $D$ of the holographic vortex becomes large at low temperature reflecting the core dissipation, and it becomes large compared to $\rho_s k$. The low temperature power dissipation $\sim 1/D$ explains why it decreases with decreasing temperature. However, the "cold atom" GP outcome is that $D$ is always small compared to $\rho_s k$. Further, counterintuitively, the $P$ diminished at low temperature; now, it was because of $P \sim D$ instead!

\begin{figure*}[t]
\includegraphics[trim=0cm 8cm 0cm 7.3cm, clip=true, scale=0.6, angle=0]{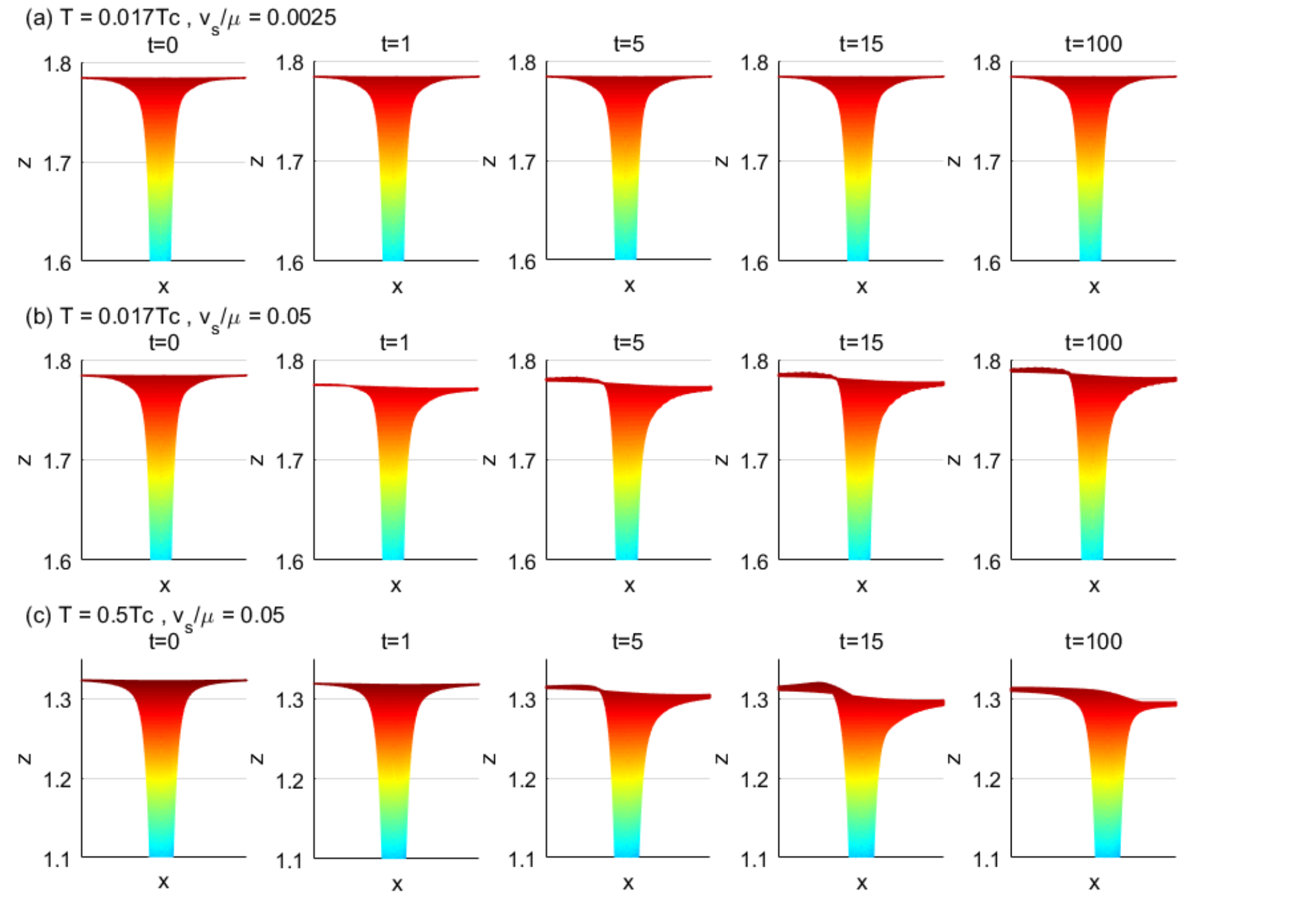}
\caption{Deformation of the vortex as function of time after the external superflow is switched on. We show the vortex in "side view" parallel to the direction of motion towards the right, and pick certain representative short times as well as the way the deformation appears in the long time ($t = 100/\mu$) stationary regime.
(a): Even at very low temperatures the vortex behaves as a rigid "particle" when the drive is small, as in Section (\ref{dissipation}).
(b,c) Both at low- and higher temperatures, the vortex starts to deform for relatively large drives. Immediately after the drive is switched on, the vortex starts to accelerate as a rigid object and this does not influence the mass as estimated in Section (\ref{massdet}). However, soon thereafter, a transient regime is followed, where the deformation develops to settle in a final "speedboat" shape in the stationary regime.
 }\label{deformationprocess}
\end{figure*}

Let us now return to the issue of core- versus the non
local dissipation associated with the normal fluid. As is
obvious from Fig. (\ref{move}), we can discriminate a core dissipation associated with the energy flux in the core region (black circle) from the "other" contributions outside the core region. The temperature evolution of these two components is also indicated in Fig. (\ref{dissipation}). According to the expectations, at low temperatures, it was completely dominated by the shear drag associated with the strange metallic core. Whereas, with increase in the temperature, the non-core contributions gradually took over the power dissipation.

\subsection{The Iordanskii force $D'$ according to holography.}
\label{Iordanskiiforce}

Let us now turn to the Iordanskii force parameter $D'$. This was introduced by Iordanskii \cite{Iordanskii} based on intuition rooted in the two fluid phenomenology, where it was insisted upon that at finite temperature, there exist a normal fluid component next to the superfluid. When dealing with the circulating superflow of the vortex, it may be naively expected that the {\em normal} fluid also exerts a Magnus force of the vortex motion. This is implicitly hard wired in the GP modelling with its simple damping parameter $\gamma$ encoding for the thermal physics. Accordingly, although smaller by an order of magnitude, the GP $D'$ parameter is according to the trajectories tracking the temperature dependence of the $D$ parameter (Fig. \ref{smallvsD} ). Both are interpreted as following the temperature dependence of the normal density $\rho_n (T)$.

However, upon closer inspection, it became clear in the course of time, that this is actually a subtle affair. Conflicting claims appeared in the literature even in the "simple" kinetic gas context. In the 1990's, Thouless et al. \cite{Thouless2} argued that $D'$ should vanish based on seemingly first principal arguments. Several results based on explicit calculations appeared with different outcomes \cite{Thompsonthesis}. This remains an unresolved problem.

What does holography indicate regarding this affair? According to the second line of Eq. (\ref{HVIsteadyvel}), we can determine $D'$ with very high sensitivity as it is directly proportional to $v_{L_y}$. This velocity component was found to be very small compared to $v_{L_x}$ determining $D$; however, it was surely still finite. The result is shown in Fig. (\ref{smallvsD} ): $D'$ is in magnitude only $\simeq 1$ \% of $D$! This suggests that in first instance, there may be truth in the claim by Thouless et al.; that is, it {\em nearly}, but not totally vanished. The difference may be in subtle high order processes that may be hard to identify. We leave this as a challenge for future work.

\section{Driving it hard: the deformation of the vortex.}
\label{Sectdeformation}

In the last section we focused purposefully on the small velocity regime characterized by a weak Magnus force, such that the system stays close to equilibrium. The question that arises is whether there are new phenomena that can be identified while ramping up $v_s$ such that the Magnus force is no longer small compared to the equilibrium scales of the system. We have identified in the holographic simulations a phenomenon that has to the best of our understanding not been observed before: when the vortex was accelerated it gradually {\em deformed}. This significantly affected the dissipative properties of the vortex motion, but only so in the low temperature local pair like regime. This study presents here no more than a descriptive discussion of these phenomena, with no intention of explaining it in any quantitative depth. We suspect that such a full elucidation of this non-equilibrium physics may be quite involved and we leave this to future work.

\subsection{ The vortex as a speedboat.}
\label{vortexwake}

While running the simulations, a phenomenon that appears to be hitherto unidentified was discovered. This is a testimony of the remarkable flexibility and precision of the holographic simulations. As explained in the previous Section, the measure of the strength of the forcing is the ratio $v_s/v_c$. What happens when $v_s$ becomes a substantial fraction of $v_c$ and how does this depend on temperature? The outcome is in the form of an unanticipated, novel dynamics, which was prominent for large $v_s$ in the low temperature regime.

\begin{figure}[h]
\includegraphics[trim=1cm 9cm 0cm 8cm, clip=true, scale=0.45, angle=0]{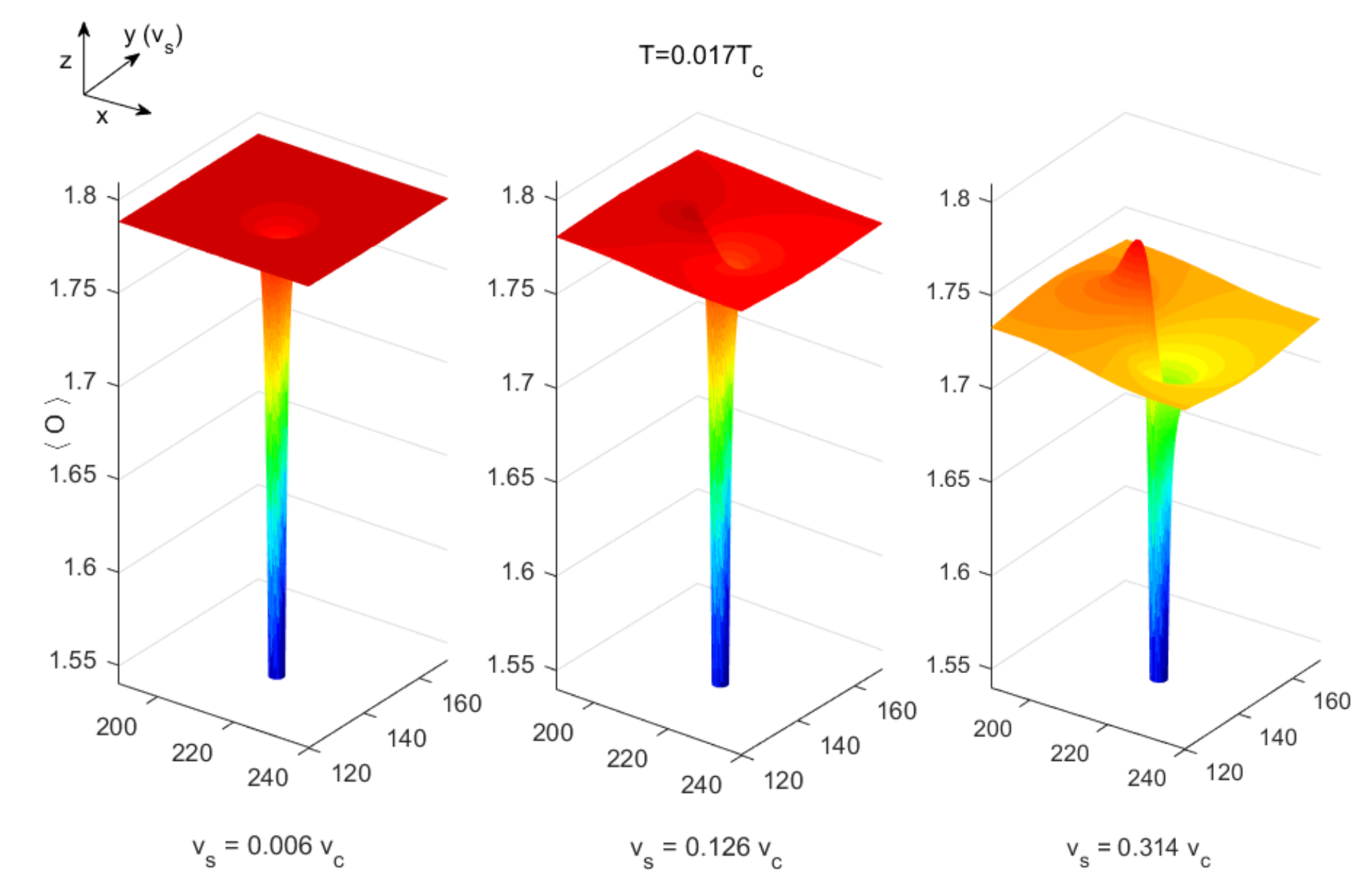}
\caption{Order parameter profile of the vortex in the stationary long time regime as function of increasing drives $v_s = 0.006v_c, 0.126v_c$, and $0.314v_c$ respectively.
For the largest current case, it is evident that the background order parameter was becoming smaller, as the system began to approach the critical current.
}\label{deformation}
\end{figure}

It works shown in Fig.(\ref{deformationprocess},\ref{deformation}). At very short times following the quench, the static vortex simply began to accelerate as a rigid body; consequently, the next aspect did not influence the estimates for the vortex mass, which was therefore implicitly defined as this "initial" mass. However, the shape of the vortex began to {\em deform} after a certain amount of time. Further, as a function of increasing $v_s$ this deformation becomes more pronounced. Figure (\ref{deformation}) shows this evolution after a time $t= 100 /\mu$, where in all cases, the deformation was itself stationary. Notice that for the largest velocity ($v_s = 0.314 v_c$), the background superfluid density was visibly reduced since this is approaching the critical velocity.

\begin{figure}[t]
\includegraphics[trim=2cm 5.3cm 2cm 5.2cm, clip=true, scale=0.55, angle=0]{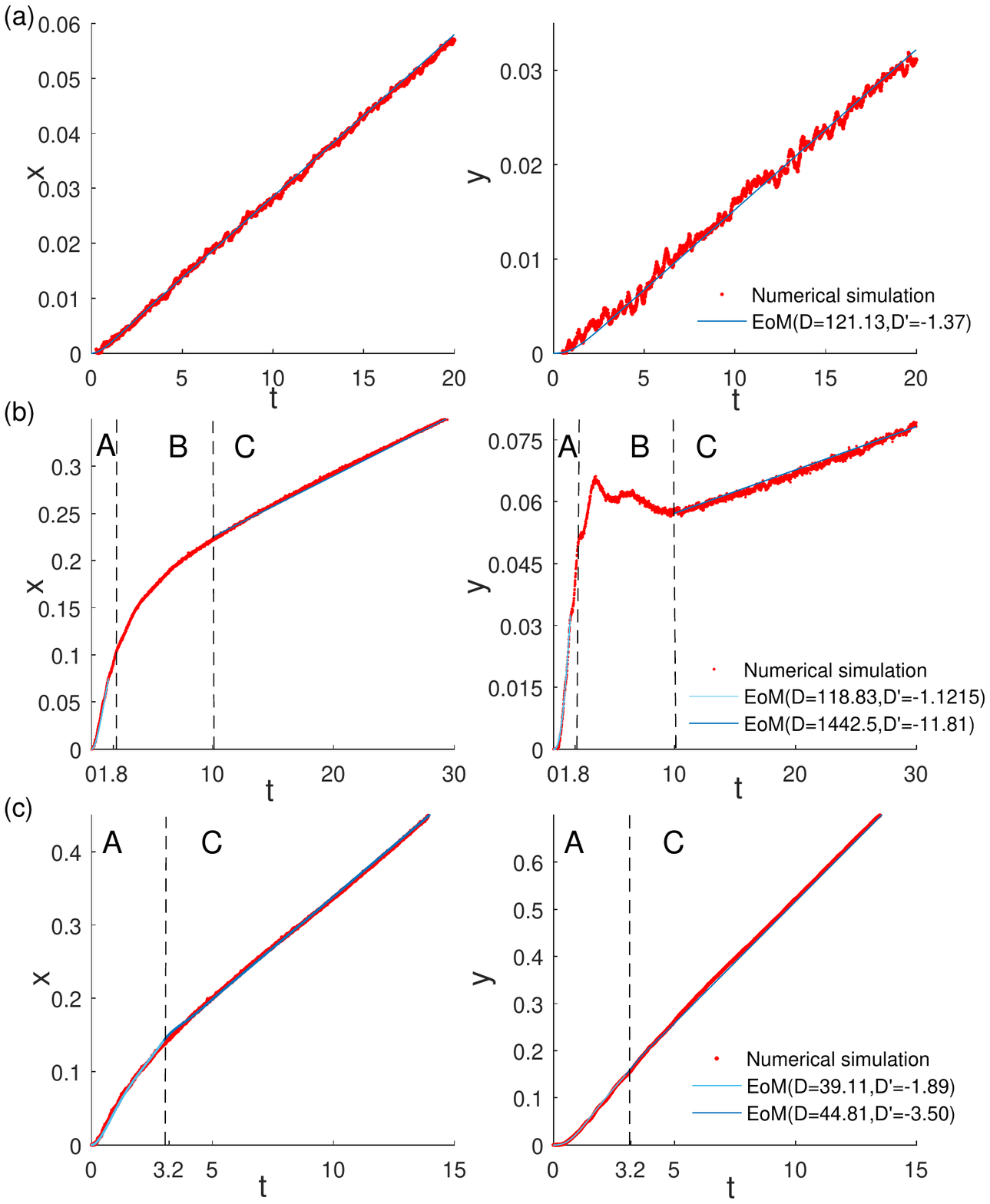}
\caption{ Dependence of the trajectories on the driving velocity $v_s$ and temperature.\\
(a) The regime of low temperature ($T=0.017T_c$) and small $v_s / \mu=0.0025$; after a very brief acceleration, the vortex only moves at a constant speed.\\
(b) The regime of low temperature ($T=0.017T_c$) and large $v_s / \mu=0.05$
where the deformation is pronounced.
At very early times (regime A), the vortex starts to accelerate in a rigid manner and the drive does not influence the vortex mass (light blue). In regime B, the deformation becomes discernible and starts to develop and fitting to the HVI equation fails. Upon entering regime C, the deformation has become stationary and the vortex again enters a drag dominated regime where, similar to the case before, the HVI coefficients $D and D'$ can be determined from the constant velocity trajectories.\\
(c) Upon raising the temperature, pronounced also for a strong drive ($v_s = 0.2 v_c$) and at $T = 0.5T_c$, the effects of the deformation became less obvious and also the coefficient change before and after deformation was not drastic.
}\label{trajectory}
\end{figure}

Similar to the previous case, the supercurrent flows in the y-direction, exerting a Magnus force in this direction on the vortex. However, for these larger velocities the (compact) vortex can no longer be considered a rigid object as its circulating currents add- and subtract to the background current at the "front" and "back" of the vortex, respectively. Consequently, the superfluid density is {\em suppressed} at the "front" and {\em enhanced} at the 'back" given that the fluid velocities despite all being of order of $v_c$ were quite different. Effectively, a "bow wave" and a "stern wake" developed in the outer parts of its core, similar as the wake of a racing speedboat!

This metaphorical similarity with the speed boat is further strengthened by observing the long time regime. The deformation requires a certain time to develop, which should be set by the inverse of the gap scale. However, at the longer times, where the vortex settles in the stationary, drag dominated the regime wherein it was moved with constant velocity (Figs. \ref{RNGP}, \ref{trajectory}). Further, the deformation itself became stationary. This is similar to the stationary wake of the speedboat when it has settled at its constant "terminal" velocity. This is feasible because it is caused by a similar balance between the drag and external forces, acting differently at the "bow" and the "stern" of the moving vortex.

\subsection{Non-linear dependence of the dissipation on the strength of the drive}
\label{deformeddiss}

As a final issue, are the dissipative parameters of the HVI EOM pending on the strength of the drive and the deformation, and if so what happens?  As we already announced, for a strong drive the vortex is deforming in a similar fashion both at low and high temperature, e.g. Fig.'s (\ref{deformationprocess}: b, c). However, we find that the degree to which {\em the trajectories} and thereby the HVI dissipative parameters {\em turn non-linear is depending critically on temperature}.

\begin{figure}[h]
\includegraphics[trim=1.8cm 11cm 0cm 10.5cm, clip=true, scale=0.55, angle=0]{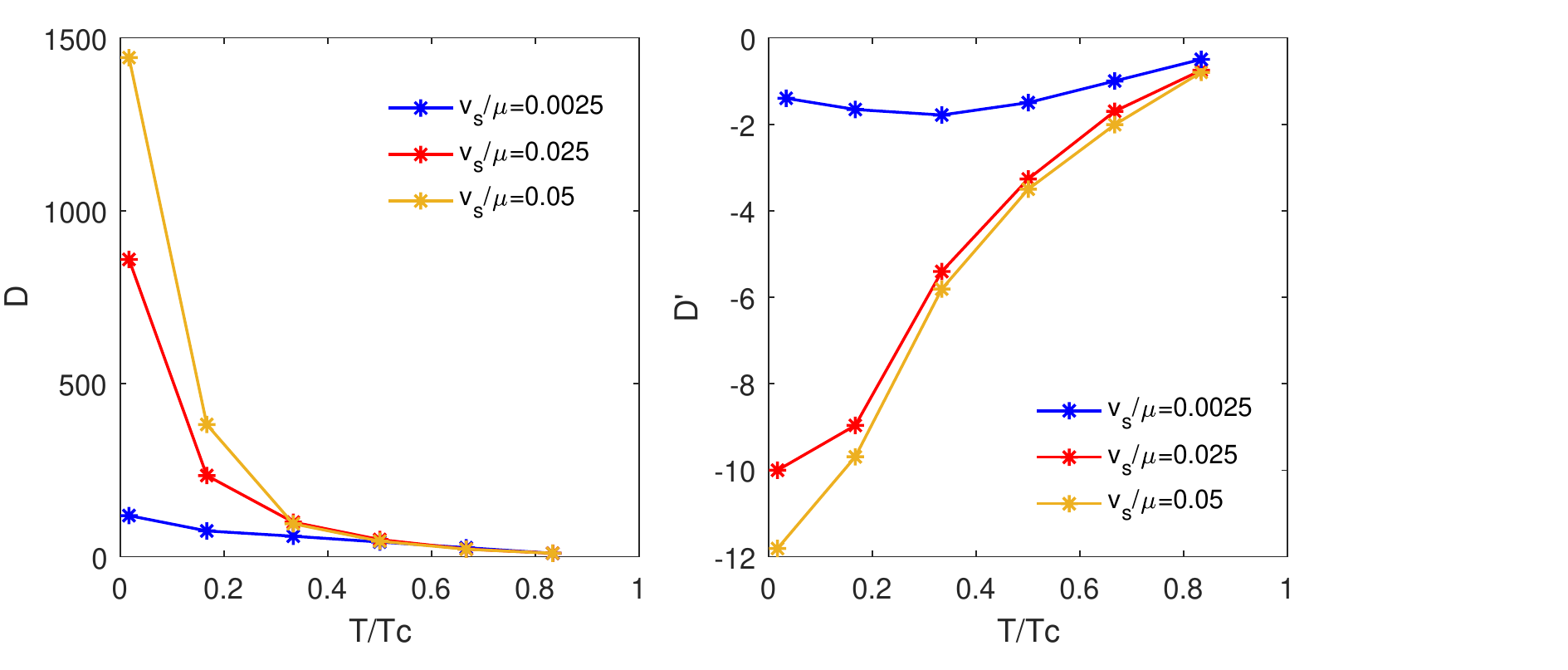}
\caption{ Coefficients $D$ and $D'$ as a function of temperature for a range of driving velocities, revealing an extreme
sensitivity to the latter at low temperatures.
}\label{DClargevs}
\end{figure}

This is manifested directly by the trajectories, as shown in Fig. (\ref{trajectory}). Panel (a) shows the small drive reference. Panel (c) shows the trajectories at $T = 0.5 T_c$ and a large $v_s = 0.2 v_c$ drive; however, these appear similar to the small drive case. Radically different trajectories were observed for large drives (shown is $v_s =0.05 \mu$) and {\em low temperature}, as in panel (b). In fact, next to the early time "mass dominated" regime A, and the late time stationary regime C, an intermediate time regime "B" was obtained, where the deformation developed. It appears that we cannot quite fit the trajectories in this regime with the HVI equation. This is not surprising as the vortex in this regime did not behave as a rigid "particle", and rather was subjected to a developing "plastic deformation" that should imprint on the dissipative properties. This is particularly obvious considering the $y$ direction (right middle panel). As we discussed, this gives away directly  the Iordanskii force; however, this exhibits erratic behavior in this deformation regime.

This can be further quantified by fitting the late time stationary trajectories to the HVI equation as before, and now also as function of the strength of the drive. The result is shown in Fig. (\ref{DClargevs}), revealing a staggering surprise! When $T \rightarrow T_c$, $v_s$ does not make any difference. It is evident that $D$ and $D'$ are the same as in the small velocity regime. However, upon lowering the temperature, $D$ and similarly $D'$ {\em increased by more than an order of magnitude} for the moderate $v_s/\mu =0.05$! For comparison we checked this for GP, which indicated that $D$ and $D'$ were completely independent of $v_s$. This is clearly an effect tied to the special nature of the holographic vortices.

\begin{figure}[h]
\includegraphics[trim=2cm 9cm 0cm 9cm, clip=true, scale=0.5, angle=0]{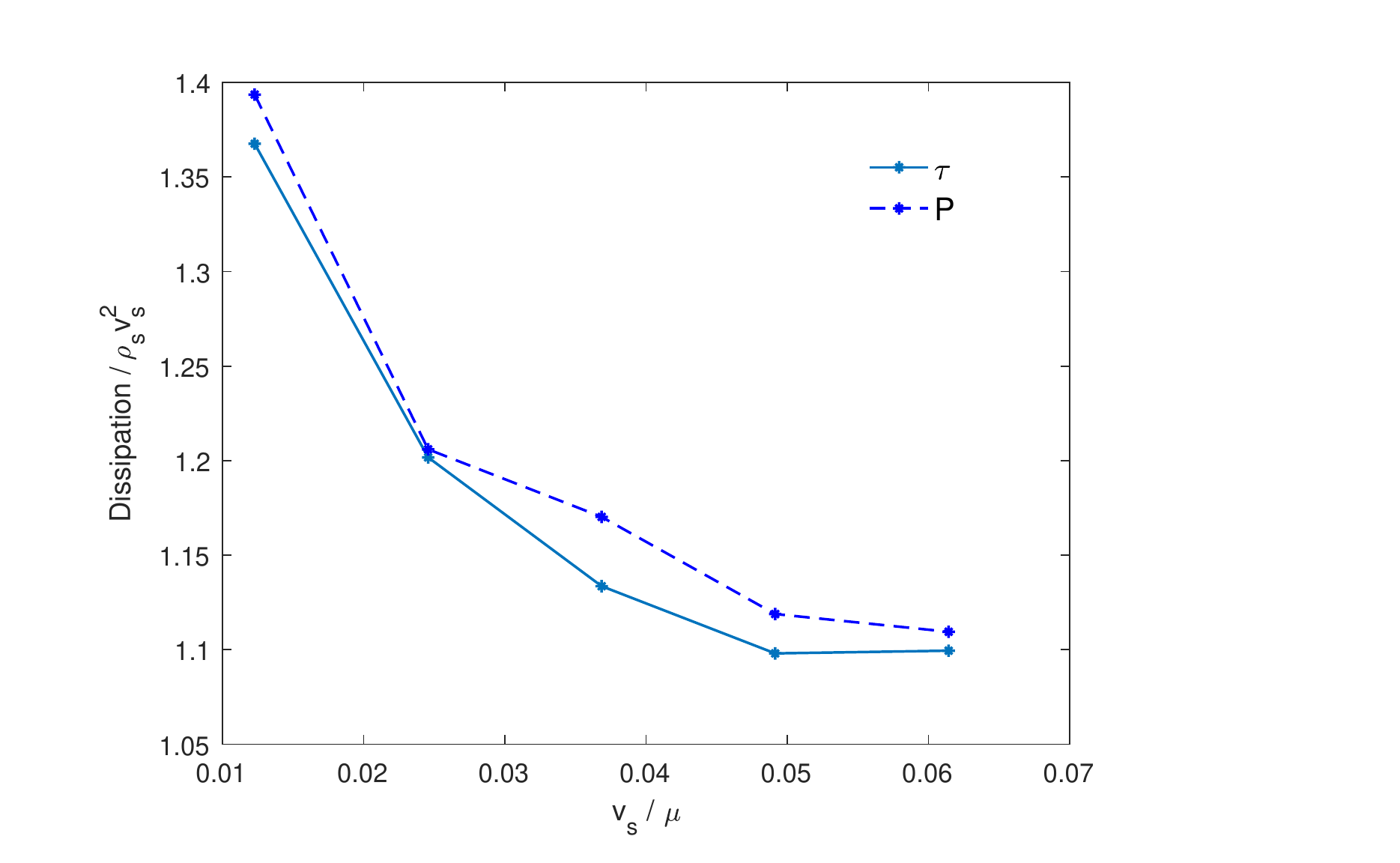}
\caption{ Relative dissipation considering the influence of deformation at different velocities. The temperature is $T=0.17T_c$. The solid line is the total energy flux from Eq. (\ref{Energyflux}), and the dashed line is the power of dissipation from Eq. (\ref{Power2}).
}\label{Dissipationvs}
\end{figure}

To shed further light on the origin of this highly non-linear behaviour of these parameters, we also inspected the energy flux $\tau$ and HVI power dissipation $P$ for different $v_s$.
Figures (\ref{Dissipationvs}) shows that with the increase in external flow velocity, the additional dragging force from deformation resulted in to a decrease in the relative dissipation, and conformed to the power of dissipation as in Eq. (\ref{Power2}).

This spectacular non-linear response of the vortex as manifested by the amplification of $D$ for increasing drive was presented as an observation. The cause remains unclear. On the one hand, one anticipates that the deformation process itself could be a source of extra dissipation; however, we observe the large increase in $D$ in the regime resulted in the deformation becoming stationary. In addition, the simple shear drag associated with the vortex core in the small drive regime, as discussed in Section (\ref{Dshear}), does not offer an obvious explanation. We do not expect that the (minimal) shear viscosity of the normal core would be affected by the drive, while the overall size of the vortex did not change significantly. Thus, this could play a role as it has the wrong sign because the vortex radius {\em increases} for increasing $v_s$. We suspect that this rather spectacular non-linear response of $D$ with $v_s$ originates from the rather anisotropic distribution of the currents implied by the "speed boat" deformation. This is an interesting subject for further research.

\section{Discussion and conclusions.}
\label{discussion}

What did we achieve? In its applications to condensed
matter physics, holography has now quite a track record
in describing an alternate reality that may not be liter-
ally realized in the laboratories, being however remark-
ably precise and correct with regard to the description of
the physical properties. A key element is its capacity to capture with high precision the ramifications of quantum thermalization, clearly indicating the macroscopic reality at finite temperature. This was previously a terrain occupied exclusively by the comparatively simple quantum kinetic theory. One virtue of holography is that in this regard it broadens the view, yielding counter examples of behaviors that are rather entirely different from this established "gas" paradigm. This is surely a beneficial circumstance dealing with problems such as those encountered in high $T_c$ superconductors, thereby offering a different basic perspective on the rather mysterious physics in these systems.

A case in point is the thermal physics of dynamical vortex systems. This is among the hardest quantum thermalization/dissipation problems that have been identified. As we reviewed in the study, even basic properties such as the vortex mass and the Iordanskii force which are within the confines of quantum kinetic gas theory are still considered as unsettled despite a large body of work.

This study showed here that with comparatively little effort we managed to complete chart this problem based on holography, now in the setting of an extremely strongly interacting and densely many body entangled underlying quantum physics. This revealed several surprises that we rationalized to a certain degree. (a) The vortex inertial mass varying from the relativistic generalization of the Baym-Chandler mass at low temperature, to subsequently vanish upon approaching the critical temperature. (b) The HVI drag coefficient ($D$) peaking at low temperature, which was associated with the particulars of the shear drag rooted in the "strange" metal core, supplemented by a subdominant normal fluid drag similar to that in the kinetic gas systems. (c) A Iordanskii force that is finite but very small and roughly temperature independent, hinted at a generalization of the arguments of Thouless et al. (d) The most genuine of all surprises in the form of the "speedboat" deformation of the vortex under a strong drive, causing a spectacular increase in the drag forces.

Is there more to it than just an entertaining theoret-
ical toy model exercise? Can this be used to shed light
on real physics problems? A first area where this may
be employed may be quantum turbulence. We focused on the most basic dynamical question: consider one vortex and drive it by a homogeneous background flow. This was input for the next level problem: consider a closed vortex loop in 3 dimensions. The counter circulating (vortex, anti-vortex) parts of the loop exerted attractive Magnus forces on each other, causing the loop to shrink and eventually annihilate. Similarly, vortex strands may intersect and reconnect. This was already studied holographically \cite{Wittmer2} although only for the case of the minimal holographic superconductivity that is characterized by a different dissipation. The cores are here associated with the zero density CFT's with viscosities that are vanishing at zero temperature -- single vortex properties are actually not even charted for these set ups but we can anticipate the outcome based on what is known.

Thus, by choosing different holographic set ups, a variety of typical single vortex dissipative behaviors can be hardwired, which can then be used to investigate the manner in which it influenced these more intricate and complex behaviors. This becomes particularly interesting dealing with full-fledged quantum turbulence, where one departs from a dense tangle of vortices to track how this evolves dynamically. This is a main stream research area that started in the Helium, being intensely pursued presently in the cold atom laboratories. The theoretical side is however nearly completely monopolized by studies departing from the GP equation, a notable exception being the study based on minimal holographic superconductivity demonstrating the "reversal" of the (Kolmogorov) cascade in 2D superfluid turbulence \cite{Chesler}. However, manner in which this is influenced by the, in principle, differing dissipative properties of single vortices depending the underlying quantum system remains unclear. Other questions are the manner in which the peculiar behaviour of the drag parameter of our RN vortices influenced the cascades, whether the vortex mass matters, and can the deformation phenomenon influence the reconnection physics.

These issues are addressing theoretical curiosity, but could these relate to circumstances found in nature? The crucial aspect of our holographic set up is the dissipative nature of the vortex core. In fact, this is an ubiquitous condition in {\em metallic} BCS superconductors. In principle, the mean field gap is suppressed in the core of the vortex of such superconductors. However, in the extreme "clean" $^3$He superfluid, the finite size splitting of the Bogoliubov type quantum mechanical states in the core become eventually large compared to temperature. This rendered these core as effectively empty, thus making it possible to identify the "anomaly currents" contributing to the dissipation associated with the unconventional order parameter \cite{Volovik}.

However, a next caveat is that this works differently in typical metallic superconductors. Invariably, there is a residual quenched disorder rendering a residual normal state resistivity owing to elastic scattering in the low temperature metallic state. As reported by Bardeen and Stephen \cite{Bardeen} this diminished the effect of the finite size gaps when the elastic mean free path was shorter than the coherence length, thus rendering the vortex cores as dissipative. Although the physics governing the dissipation from our RN holographic vortices is fundamentally different (the hydrodynamical shear drag), there may be sufficient similarity for holographic simulations to be useful to study dynamical questions in this context.

There is yet another difference of principle. Metallic superconductors were {\em gauged} and this rendered the physics to be different from the neutral superfluid described using our present set up. This is particularly the case for the topological excitations, which are now "fuxoids" (Abrikosov-, Nielsen-Olesen vortices), characterized by an absolute size set by the London penetration depth being characterized by the magnetic flux quantum as topological charge. These are sourced by magnetic fields instead of "mechanical" external forces. They played no role at all in the study of quantum turbulence for the simple practical reason that the typical time scales associated with ramping up magnetic fields were very long compared to the intrinsic time scales associated with the (electronic) fluxoids. Consequently, it was impossible to quench the system, which is a prerequisite to reach far out of equilibrium conditions.

On the other hand, motivated by applications "vortex dynamics" involving issue like flux penetration, the establishment and pinning of Abrikosov (fluxoid) lattices have been subject of an intense research effort \cite{Blatter}. But also in this context realistic simulations of such dynamical phenomena is not easy given the intricacies of vortex dissipation.

In this regard, very recently it got completely clarified how to use the holographic dictionary to "lift" the neutral superfluid in the holographic boundary to a gauged status in an technically efficient manner, in terms of mixed (direct and alternate) boundary conditions \cite{Ahn,Montull,Domenech}. It is straightforward to incorporate these in the general framework that we have been using here, and it would be quite interesting to explore what can be learned from holography in the context of this large research area dealing with magnetic fields in superconductors.

\section*{Acknowledge}

H.B. Z. acknowledges the support by the National Natural Science Foundation of China (under Grants No. 11275233) and M. T. acknowledges the support by the
JSPS KAKENHI (under Grant No. JP20H01855).


%

\end{document}